\documentclass[aps,prl,amsmath,twocolumn,amssymb,showpacs,titlepage,showkeys,superscriptaddress]{revtex4-1}

\usepackage{colortbl}

\usepackage{graphicx}
\usepackage{nicefrac}
\usepackage{amsfonts}

\usepackage{amssymb}
\usepackage{amsmath} 
\usepackage{bm}

\usepackage{subfigure}
\usepackage{multirow} 
\usepackage{tabularx} 
\usepackage{array}

\usepackage{units}

\usepackage{tensor} 
\usepackage{braket}

\usepackage{bm}
\usepackage{hyperref}

\newcolumntype{?}{!{\vrule width 0.8pt}}

\usepackage{color}
\usepackage[usenames,dvipsnames]{xcolor}
\usepackage{makecell}

\begin{document}

\title{Osmates on the verge of a Hund's-Mott transition: \\ The different fates of NaOsO$_3$ and LiOsO$_3$}
\author{Daniel Springer}
\affiliation{Institute of Solid State Physics, TU Wien, A-1040 Vienna, Austria}
\affiliation{Institute of Advanced Research in Artificial Intelligence, IARAI, A-1030 Vienna, Austria}
\author{Bongjae Kim}
\affiliation{Department of Physics, Kunsan National University, Gunsan 54150, Korea}
\author{Peitao Liu}
\affiliation{University of Vienna, Faculty of Physics and Center for Computational Materials Science, Vienna, Austria}
\author{Sergii Khmelevskyi}
\affiliation{Research Center for Materials Science and Enginireeng, TU Wien, A-1040 Vienna, Austria}
\author{Severino Adler}
\affiliation{Institute of Solid State Physics, TU Wien, A-1040 Vienna, Austria}
\affiliation{Institut f\"ur Theoretische Physik und Astrophysik and W\"urzburg-Dresden Cluster of Excellence ct.qmat, Universit\"at W\"urzburg, 97074 W\"urzburg, Germany}
\author{\\ Massimo Capone}
\affiliation{CNR-IOM-Democritos National Simulation Centre and International School for Advanced Studies (SISSA), Via Bonomea
265, I-34136 Trieste, Italy}
\author{Giorgio Sangiovanni}
\affiliation{Institut f\"ur Theoretische Physik und Astrophysik and W\"urzburg-Dresden Cluster of Excellence ct.qmat, Universit\"at W\"urzburg, 97074 W\"urzburg, Germany}
\author{Cesare Franchini}
\affiliation{University of Vienna, Faculty of Physics and Center for Computational Materials Science, Vienna, Austria}
\affiliation{Dipartimento di Fisica e Astronomia, Universit\`{a} di Bologna, 40127 
Bologna, Italy}
\author{Alessandro Toschi}
\affiliation{Institute of  Solid State Physics, TU Wien, A-1040 Vienna, Austria}
\begin{abstract}
We clarify the origin of the strikingly different spectroscopic properties of the chemically similar compounds NaOsO$_3$ and LiOsO$_3$. Our first-principle many-body analysis demonstrates that the highly sensitive physics of these two materials is controlled by their proximity to an adjacent Hund's-Mott insulating phase. Although $5d$ oxides are mildly correlated, we show that the cooperative action of intraorbital repulsion and Hund's exchange becomes the dominant physical mechanism in these materials, if their $t_{2g}$-shell is half-filled.
Small material-specific details hence result in an extremely sharp change of the electronic mobility, explaining the surprisingly different properties of the paramagnetic high-temperature phases of the two compounds.
\end{abstract}
\maketitle

Metal-insulator transitions driven by the Coulomb repulsion (Mott transitions) represent one of the most characteristic hallmarks of electronic correlations.
A variety of  Mott transitions \cite{RevModPhys.70.1039} are observed in $3d$ transition metal oxides (TMO), among which we recall the prototypical case of V$_2$O$_3$\cite{PhysRevB.7.1920, RevModPhys.70.1039, pssb.201248476}. Yet, when considering TMOs with heavier transition-metal elements, the impact of correlations gets weakened due to the larger spatial extension of the $4d$ and $5d$ electronic orbitals. This may cause correlation effects to become comparable to other physical mechanisms at work, such as spin orbit coupling (SOC)\cite{PhysRevLett.102.017205, JPhysCM.29.263001, PhysRevLett.107.266404} and Stoner magnetism.

Aside from specific situations, where an enhancement of correlations can be triggered by the realization of effective single-orbital configuration \cite{PhysRevLett.101.076402, PhysRevLett.107.266404, PhysRevB.92.054428, PhysRevB.95.115111, NatComm.8.14407, Nature.554.341}, one expects systematically weaker many-body effects in $5d$ TMOs. One should then observe just mildly renormalized Fermi-liquid properties in these materials, provided that the temperature is high enough to destroy any possible long-range order. However, if a half-filled configuration between nearly degenerate orbitals is realized (e.g.\ $3$ electrons in the $t_{2g}$ orbitals), $4d$         \cite{PhysRevLett.108.197202} and 
{\sl even} $5d$ materials may display significant correlation effects.
We showcase that the latter scenario is not only an academic problem by demonstrating its actual realization in the elusive physics of two osmates, NaOsO$_3$ and LiOsO$_3$.

We recall that in general, the presence of $n\! = \! 3$ electrons in the three $t_{2g}$ orbitals can trigger the formation of very stable high-spin ground states, even if the intra-orbital local Coulomb repulsion is not particularly large. If this happens,  the {\sl localizing} strength of the on-site Hund's exchange coupling ($J$), favoring the onset of a Hund's-Mott insulating state, gets significantly magnified.
This trend is opposite to the extensively studied case of systems with non half-filled $t_{2g}$ orbitals (e.g. $n\! = \! 2, 4$), where $J$ has a {\sl delocalizing} effect for the ground-state properties 
\cite{PhysRevLett.107.256401, AnnuRevCMP83}.
The tendency towards the formation of so-called Hund's {\sl metals} at $n\!= \!2$ can be ascribed to the competition between two possible localized states \cite{PhysRevLett.122.186401}: a pure Mott state where the double occupancy is minimized and a charge-disproportionated Hund's insulator, where local spin moment is maximized. 

Instead, for the case $n \! = \! 3$, discussed in the present work, the intra-orbital Coulomb repulsion {\sl and} the Hund's coupling favor the same insulating state: the Hund's-Mott insulator, simultaneously associated to a high spin configuration {\sl and} suppressed charge fluctuations. 
Thus, in the case of a nominally half-filled $t_{2g}$ configuration, $U$ {\sl and} $J$ act cooperatively to suppress the electronic mobility. 
It is this {\sl synergic} action of $U$ and $J$, which strongly lowers the critical interaction values for the onset of Hund's-Mott phases and hence allows the associated physics to emerge even in less correlated $5d$ compounds.
At the same time, this explains why correlation effects fade out extremely quickly, when only slightly moving away from
the Hund's-Mott insulating region ($n\!=\! 3$ in the $t_{2g}$ orbitals) in contrast to the much smoother 
evolution observed in the Hund's metals \footnote{Cf. the general analysis of Refs.~\onlinecite{PhysRevLett.107.256401, ldmmcchapter}}.

To illustrate the main physical trends, we present in Fig.~1 a {\sl generalized version} of the paramagnetic phase-diagram of TMOs, extended to {\sl also} include the $5d$ compounds. The formation of a local moment, driven by the on-site interactions, stabilizes the corresponding Mott insulating phases (grey-shadowed zones) for different fillings (n\!=\!2, 3) of the t$_{2g}$ orbitals. Evidently, the most favorable conditions to induce or, at least, to approach a Mott phase in the less correlated $5d$ TMOs are offered by the half-filled $t_{2g}$ configuration. 


As the Hund's-Mott phase is characterized by a high sensitivity with respect to small perturbations (e.g. lifting of the degeneracy of the $t_{2g}$ orbitals, hybridization with the $p$ orbitals of the ligands, etc.), even tiny differences in realistic systems may quench correlation effects and strongly modify the physical properties.
The pair of (nominally half-filled) osmates, NaOsO$_3$ and LiOsO$_3$, offers a particularly promising playground for observing genuine Hund's-Mott-driven physics in otherwise mildly correlated 5d oxides. In spite of  a similar chemical composition, the two compounds display quite different physical properties, even in their high-temperature regimes.

In particular, for the orthorhombic perovskite NaOsO$_3$ terahertz and infrared spectroscopy \cite{LoVecchio2013} have revealed the presence of a high temperature ($T$) metallic state with relatively good Fermi liquid properties. Upon cooling, a MIT is accompanied by the emergence of a long-range ($G-$type) antiferromagnetic order at $T\!=\!T_N \sim 410$K \cite{PhysRevB.80.161104}. The physics below $T_N$ initially was interpreted as that of the elusive Slater insulating state\cite{PhysRevLett.108.257209} and later  better defined in terms of a spin-fluctuation-driven Lifshitz transition \cite{PhysRevB.94.241113, PhysRevLett.120.227203, PhysRevB.97.184429}. Hence both, the high-$T$ metallic phase of NaOsO$_3$ and its antiferromagnetic behavior below $T_N$, appear compatible with a weakly correlated scenario.

In contrast, the rhombohedral LiOsO$_3$ does \emph{not} exhibit any magnetic ordering at low-$T$, but instead features a phase-transition to a non-centrosymmetric structure for $T < T_c \!=\!140$ K\cite{NatMat.12.1024}.  The essential difference emerging in comparison to NaOsO$_3$ is, that upon raising the temperature, the optical spectra of LiOsO$_3$ very rapidly lose any sign of metallic coherence\cite{PhysRevB.93.161113}. The Drude peak is even replaced by a slight low-frequency {\sl downturn} already at $T\!=\!300$ K.
In fact, the temperature trend of the infrared spectra observed for LiOsO$_3$ surprisingly well resembles that of undoped V$_2$O$_3$ \cite{PhysRevB.77.113107}, the prototypical $3d$ TMO on the verge of a Mott MIT\cite{PhysRevB.7.1920, RevModPhys.70.1039, pssb.201248476}.

\begin{figure}[t]
\centering
 \includegraphics[width=0.47\textwidth]{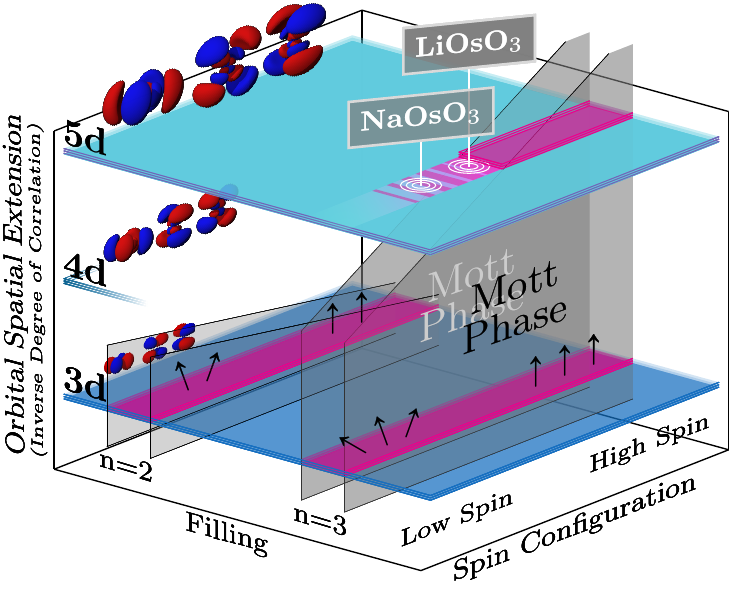}
 \caption{(Color online) Schematic representation of the Mott-insulating phases (magenta stripes along the horizontal planes, the shadowed grey vertical planes are guides to the eye) in the parameter space of a generic TMO with correlated orbitals of different spatial extension (increasing from $3d$ to $4d$ and $5d$). The picture distinguishes the effects of high/low spin configurations on the insulating phases at different fillings ($n\!=\!2$ and $n\!=\!3$). The supposed location of the pair of osmates NaOsO$_3$ and LiOsO$_3$ in the highly sensitive parameter region in immediate proximity to the $n\!=\! 3$ Hund's-Mott metal-insulator transition is indicated by white concentric circles.}
 \label{Fig:PhaseDiagram}
\end{figure}

Given the very similar energy scales (width of the conduction bands, size of the screened electronic interaction) and the (half-filled) electronic configuration of the two materials, one of the few plausible explanations for the sharp contrast between the high-$T$ spectral properties of NaOsO$_3$ and LiOsO$_3$ would be their proximity to a Hund's-Mott insulating phase as depicted in Fig.~\ref{Fig:PhaseDiagram}.

\begin{figure*}[t]
\centering
 \includegraphics[width=0.49\textwidth]{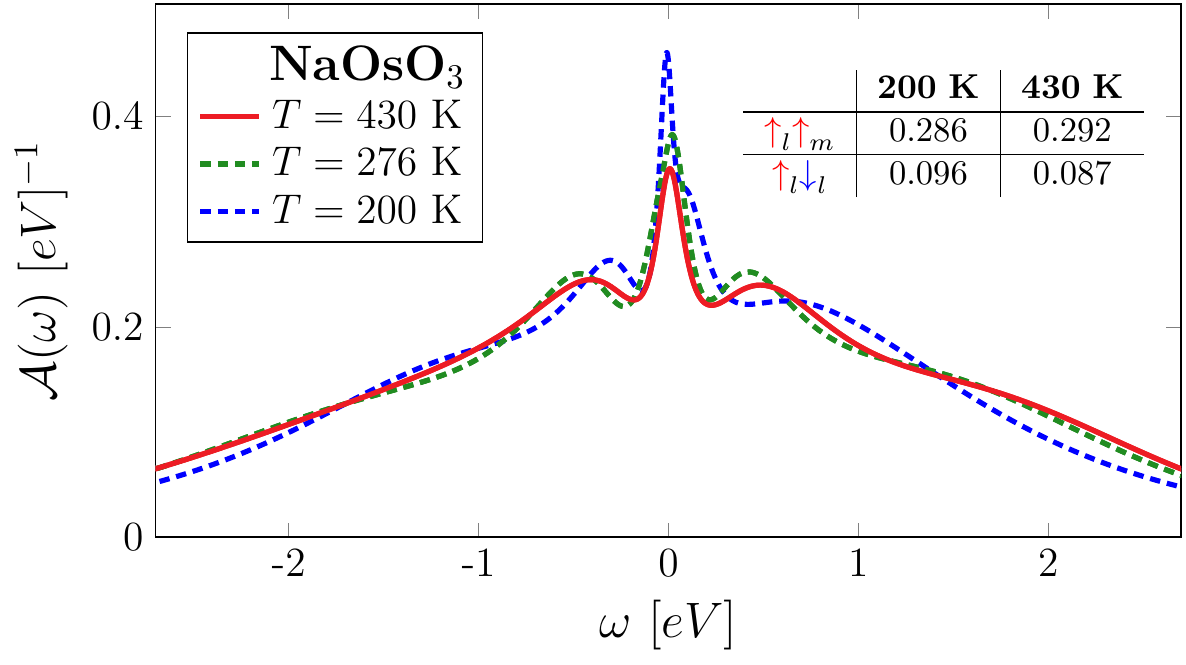}
 \includegraphics[width=0.49\textwidth]{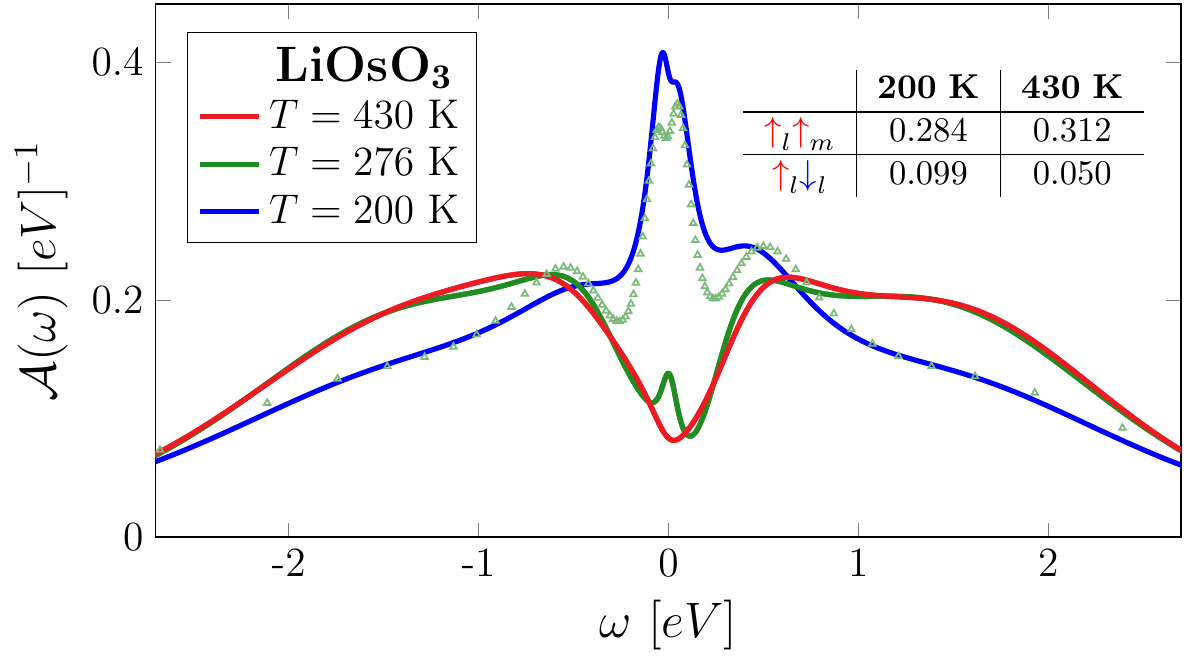}
 \caption{(Color online) Temperature evolution of the on-site spectral function in the {\sl paramagnetic phases} of NaOsO$_3$ (left) and LiOsO$_3$ (right) computed with DFT+DMFT on the $t_{2g}$-orbitals of Os (averaged over the Os atoms in the unit cell~\cite{Suppl}). 
 Dashed lines (left panel) refer to paramagnetic calculations below the $T_N$ of NaOsO$_3$ (where the experimentally observed gap is associated to the onset of a magnetic order\cite{LoVecchio2013}), while the empty triangles (right panel) mark the most metallic of the two stable DMFT solutions of LiOsO$_3$ found at $T\!=\!276$ K.  {\sl Inset}: Values of the averaged intra- and inter-orbital double-occupancy at different $T$.}
\label{Fig:FullCoulomb}
\end{figure*}

To demonstrate that this scenario indeed applies, we perform state-of-the-art {\sl ab-initio} many-body calculations  of the {\sl paramagnetic} phases for both compounds on an equal footing, exploiting the density functional theory (DFT)\cite{Kresse1993,Kresse1996} merged with dynamical mean field theory (DMFT)\cite{RevModPhys.68.13}. 
The DMFT calculations have been performed for the Wannier-projected  $t_{2g}$ orbitals of Os\footnote{In our maximally localized Wannier projection we have effectively downfolded both the high-lying $e_g$ orbitals of Os, which are located approximately $2$ eV, as well as the low-lying $2p$ orbitals of the O~\cite{Suppl}} considering the full Coulomb (screened) interaction on Os, calculated by means of the constrained random phase approximation (cRPA) for both compounds~\cite{Suppl} in their respective paramagnetic phases. Aside from specific bandstructure details~\cite{Suppl,Peitao2020}, the main energy scales in the two materials are quite similar.

The $t_{2g}$ bandwidths are $3.9$~eV (NaOsO$_3$) and $3.5$~eV (LiOsO$_3$); the local orbital splittings of the $t_{2g}$ manifold are about $150$~meV and $250$~meV, while the averaged intra-orbital Coulomb repulsion ${U}$ is $2.25$~eV and $2.35$~eV and the averaged local Hund's exchange coupling ${J}$ is $0.24$~eV and $0.25$~eV for NaOsO$_3$ and LiOsO$_3$ respectively~\cite{Suppl,Peitao2020}. Additionally we examined the role of SOC~\cite{Suppl}, finding that an estimated SOC of the order of $ \sim 0.3$ eV\cite{Suppl,Peitao2020} only weakly affects the splitting of the $t_{2g}$ manifold in agreement with experimental findings~\cite{PhysRevB.95.020413}. The corresponding minor quantitative changes are not significant for our study of the paramagnetic phases \cite{Suppl}.  
A larger impact of SOC may be expected in the magnetically-ordered phases of $5d^3$ compounds~\cite{Calder2016, PhysRevB.93.220408, PhysRevB.91.075133, PhysRevB.95.020413}, whose investigation is not the subject of our work.

The corresponding multi-orbital Hamiltonian, which we solved in DMFT by means of the {\sl w2dynamics} package\cite{Comput.Phys.Commun.235.388}, reads

\begin{eqnarray}
H &=& \sum_{{\bf k} \sigma lm} \,  H_{lm}^{\phantom \dagger}({\bf k}) \; c^{\dagger}_{{\bf k} l \sigma}\, c^{\phantom \dagger}_{{\bf k} m \sigma} \nonumber \\
&+& \sum_{{\bf r} \sigma \sigma'}\sum_{lmno} \, U^{\phantom \dagger}_{lmno} \, c^\dagger_{{\bf r} l \sigma}\, 
c^{\phantom \dagger}_{{\bf r} m \sigma} c^\dagger_{{\bf r} n \sigma'}\, c^{\phantom \dagger}_{{\bf r} o \sigma'}
\label{Eq:Hamiltonian}
\end{eqnarray}
where $l,m,n,o$ are the $t_{2g}$ orbital indices, $\boldsymbol{k}$ denotes the fermionic momentum, $\boldsymbol{r}$ the lattice site and $\sigma,\sigma'$ the spin. Here $H_{lm}({\bf k})$ and $U_{lmno}$ represent the single particle part and the local interaction of the low-energy Hamiltonian for the subspace of the $t_{2g}$ Os orbitals.
As usual, the best quantitative agreement with experimental trends is obtained for slightly enhanced cRPA values, here about $8$\% for {\sl both} compounds \cite{Suppl}, which is within the accuracy of cRPA estimated by the more sophisticated constrained functional renormalization group procedure \cite{PhysRevB.98.155132, PhysRevB.98.235151}. The Hund's exchange values of our calculations \cite{Suppl}, in any case, remain smaller than those assumed in previous DMFT studies~\cite{PhysRevB.93.161113} of the low-$T$ behavior of LiOsO$_3$. 

The temperature dependence of the local spectral function $\mathcal{A}(\omega)$ of the two-compounds in their {\sl paramagnetic} phases, obtained from analytic continuation~\cite{LEVY2017149,GAENKO2017235,1742-5468-2011-05-P05001} of the on-site Green's function, is reported in Fig.~\ref{Fig:FullCoulomb}.
We clearly see that the NaOsO$_3$ spectrum displays a sizably renormalized coherent peak up to the highest $T\!=\!430$ K$ > T_N$ considered, where a metallic behavior clearly emerges in the spectroscopic experiments~\cite{LoVecchio2013}. At the same time, the corresponding quasi-particle peak visible in the low-$T$ regime of LiOsO$_3$ is quickly washed out upon increasing $T$, with a complete loss of coherence  {\sl already} at/above room-$T$. The coexistence of two DMFT solutions \cite{RevModPhys.68.13}, a typical hallmark \cite{pssb.201248476} 
of a close proximity to a Mott-MIT, is found for LiOsO$_3$ at the intermediate temperature of $276$ K \footnote{At the other temperatures shown n Fig.~\ref{Fig:FullCoulomb} we did not observe a coexistence of insulating and metallic DMFT solutions for LiOsO$_3$.}. 
The insets in Fig.~\ref{Fig:FullCoulomb} indicate that the consistent loss of spectral weight at high temperatures in LiOsO$_3$ is linked to an increase of the inter-orbital spin alignment ($\langle \color{red}{\bm{\uparrow}}_{\color{black}{l}}$ ${\color{red}{\bm{\uparrow}}}_{\color{black}{m}} \rangle$) and a corresponding suppression of the intra-orbital double occupancy ($\langle \color{red}{\bm{\uparrow}}_{\color{black}{l}}$ ${\color{blue}{\bm{\downarrow}}}_{\color{black}{l}} \rangle$) \cite{Suppl}.

Hence, our DMFT results reproduce the experimentally observed presence (disappearance) of coherent quasi-particle excitations in the {\sl high-temperature} infrared spectra of NaOsO$_3$ (LiOsO$_3$), suggesting the validity of the scenario depicted in Fig.~\ref{Fig:PhaseDiagram}.	

In order to firmly establish, that the agreement between our calculation and experimental data is indeed explained by the proximity of both osmates to the Hund's-Mott insulator, we consider a simplified interaction term, which describes in the most direct way the formation of large local magnetic moments associated with the Hund's interaction.
\begin{figure*}[t]
\centering
 \includegraphics[width=0.49\textwidth]{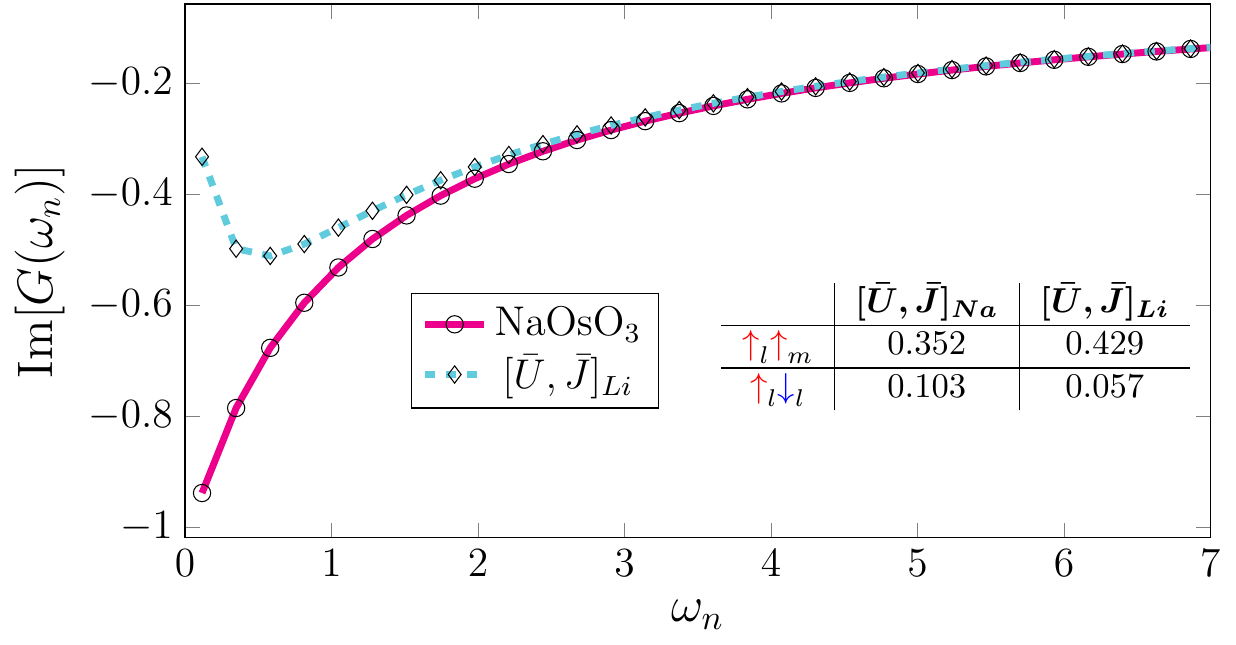}
 \includegraphics[width=0.49\textwidth]{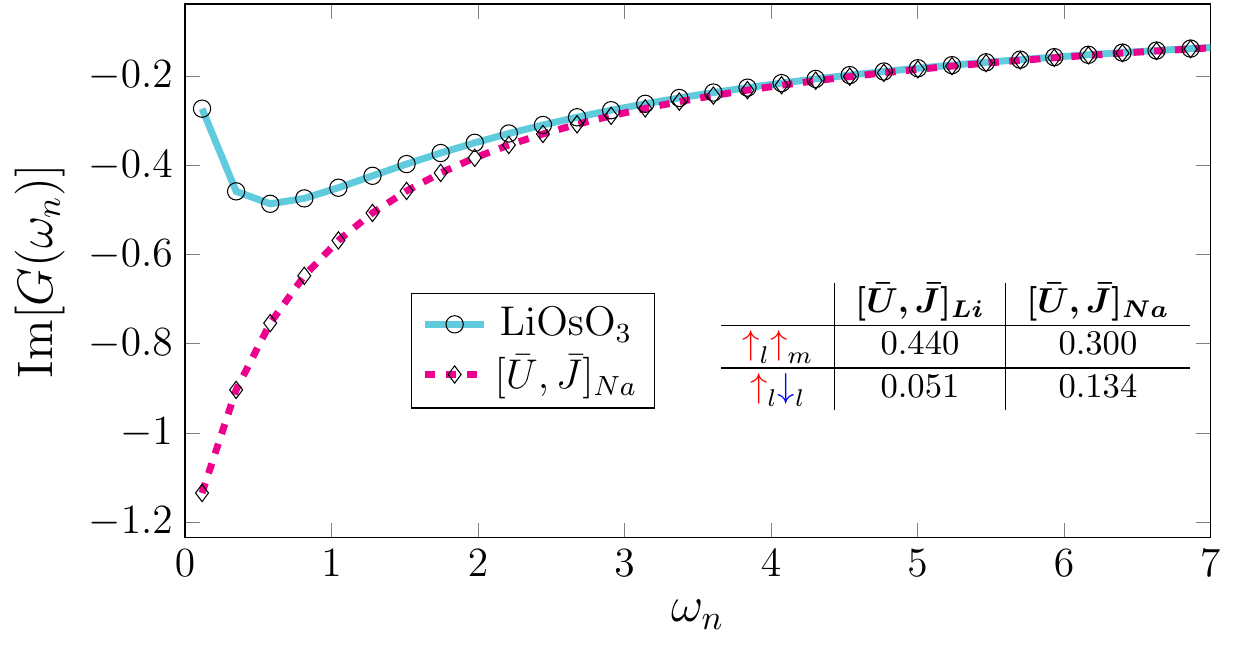}
 \caption{(Color online) DFT+DMFT calculation of the local Green's function for NaOsO$_3$ (left: solid line) and LiOsO$_3$ (right: solid line) at $T \! = \! 430$~K performed in the density-density approximation with interaction parameters ($\bar{U},\bar{J}$) \cite{Suppl}. ``Gedankenexperiment'': The dashed lines correspond to Green's functions computed using interaction values of the respective other compound. {\sl Inset}: Values of the averaged intra-orbital and inter-orbital double-occupancy at $T \! = \! 430$~K.}
\label{Fig:Gedankenexperiment}
\end{figure*}
In practice we assume an interaction of density-density character by neglecting in Eq.~\eqref{Eq:Hamiltonian} the  quantum fluctuation ($c_{{\bf r}l\uparrow}^\dagger c_{{\bf r}l\downarrow}^{\phantom{\dagger}} c_{{\bf r}m\downarrow}^\dagger c_{{\bf r}m\uparrow}^{\phantom{\dagger}}$), pair-hopping ($c_{{\bf r}l\uparrow}^\dagger c_{{\bf r}l\downarrow}^{\dagger} c_{{\bf r}m\uparrow}^{\phantom{\dagger}} c_{{\bf r}m\downarrow}^{\phantom{\dagger}}$) and correlated-hopping ($c_{{\bf r}l\uparrow}^\dagger c_{{\bf r}l\uparrow}^{\phantom{\dagger}} c_{{\bf r}m\downarrow}^\dagger c_{{\bf r}l\downarrow}^{\phantom{\dagger}}$) terms 
in the interaction as well as the orbital off-diagonal hopping terms ($c_{{\bf r}m\sigma}^\dagger c_{{\bf r}l\sigma}^{\phantom{\dagger}}$),
which are responsible for the small violation of the (local) $t_{2g}$-degeneracy, in our DMFT calculations.  As a result of these approximations, the Hund's-driven orbital off-diagonal spin correlations are enhanced, increasing the localization effects.

Our DMFT calculations (Fig.~\ref{Fig:Gedankenexperiment}) show that this modification pushes LiOsO$_3$ well inside the adjacent insulating regime, while NaOsO$_3$ remains metallic. The corresponding on-site Green's function $G$ as a function of the Matsubara frequencies i$\omega_n$ is shown in Fig.~\ref{Fig:Gedankenexperiment} for $T\!=\!430$ K.
It displays an evident metallic behavior for NaOsO$_3$, which qualitatively reproduces the calculation with the full Coulomb interaction. For LiOsO$_3$ instead, we find a vanishing Im$G(i\omega_n)$ at the Fermi edge ($\omega_n\!\rightarrow \!0$), characteristic of an insulating state. 

The Hund's-Mott nature of this insulating state can be immediately inferred from the double occupancy data of LiOsO$_3$ (inset table), which features a large local moment with almost complete inter-orbital spin alignment ($\langle \color{red}{\bm{\uparrow}}_{\color{black}{l}}$ ${\color{red}{\bm{\uparrow}}}_{\color{black}{m}} \rangle$  $\!=\!0.5$) and a simultaneous suppression of the intraorbital double occupancy. 
In NaOsO$_3$ instead, significantly smaller values of  $ \langle \color{red}{\bm{\uparrow}}_{\color{black}{l}}$ ${\color{red}{\bm{\uparrow}}}_{\color{black}{m}} \rangle $ are found, much closer to what is expected in the uncorrelated case ($\langle \color{red}{\bm{\uparrow}}_{\color{black}{l}}$ ${\color{red}{\bm{\uparrow}}}_{\color{black}{m}} \rangle$  $\!=\!0.25$).

Finally, to separate the impact of mere correlation effects from all other potentially relevant aspects (e.g. specific features in the bandstructure),
we devise a simple (numerical) {\sl "Gedankenexperiment"}, where we perform a new DMFT  (density-density) calculation for each of the two osmates, but with {\sl interchanged} interaction parameters ($\bar{U},\bar{J}$ \cite{Suppl}). The consequence of this swapping of interaction parameters is an almost perfect mirroring of the imaginary parts of the Green's functions (Fig.~\ref{Fig:Gedankenexperiment} dashed lines). The Green's function for LiOsO$_3$ with the slightly different interaction parameters of NaOsO$_3$ displays clear metallic features at the Fermi edge, similar to the original NaOsO$_3$ results. For NaOsO$_3$ on the other hand, the slightly larger interaction parameters of LiOsO$_3$ suffice to push it through the Hund's-Mott transition into the insulating regime. The values of the corresponding spin-alignment get also almost perfectly interchanged. This illustrates how the significant differences in the results of our calculations are driven by small but critical changes in the interaction parameters, convincingly confirming that the physics of the two materials is controlled by the proximity to a Hund's-Mott insulating transition.

In summary, we performed DFT+DMFT calculations for the paramagnetic phases of NaOsO$_3$ and LiOsO$_3$ with realistic, material specific values of their on-site Coulomb interaction. Our results compare remarkably well with experimental findings for the paramagnetic phases of both compounds. We are able to reproduce the occurrence of a crossover from a metallic to an incoherent behavior at room temperature, as observed in the optical spectra of LiOsO$_3$, as well as its absence in the high-$T$ spectroscopy data of NaOsO$_3$. We attribute the different spectroscopic properties of the two osmates considered to their slightly different proximity to a Hund's-Mott insulating phase.

A rather general picture emerges from our DMFT calculations and  {\sl Gedankenexperiment}: The synergic action of $U$ and $J$ for half-filled configurations is strong enough 
to induce significant correlation effects in $5d$ oxides and to drive them on the verge of Hund's-Mott MITs.
In fact, nearly half-filled  $5d$-compounds might even provide a better playground for truly observing this physics than their $3d$ or $4d$ counterparts, where it is commonly expected.  This is because a larger SOC and tetragonal/trigonal crystal-fields may successfully conspire to suppress magnetism, which often hides the major hallmarks of the proximity to Hund's-Mott insulating phases in $3d$ or $4d$ oxides\cite{PhysRevLett.108.197202,PhysRevB.95.205115,JPSJ.37.275,PhysRevB.74.144102}.  Our findings open the perspective of triggering a similar physics in other $5d$-oxides, by artificially engineering half-filled configurations in nearly degenerate $t_{2g}$-orbitals.

\begin{acknowledgments}  {\sl Acknowledgments -} We thank Luca de' Medici, Andreas Hausoel, Minjae Kim, Antoine Georges  and Jan Tomczak for interesting discussions, and the Flatiron  Institute for the hospitality in an extremely stimulating scientific atmosphere (GS, AT). We acknowledge financial support from the Austrian Science Fund (FWF) through the projects I 2794-N35 (AT) and the SFB ViCoM F41 (DS).  B.K acknowledges support from the National Research Foundation (NRF) Korea (No.2018R1D1A1A02086051),
M.C. from MIUR PRIN 2015 (Prot. 2015C5SEJJ001), MIUR PRIN 2017 "CENTRAL" (Prot. 20172H2SC4\_004) and
G.S. and S.A. from the DFG through the W\"urzburg-Dresden Cluster of Excellence on Complexity and Topology in Quantum Matter -- \textit{ct.qmat} (EXC 2147, project-ID 390858490) and through project-ID 258499086 -- SFB 1170. Calculations have been performed on the Vienna Scientific Cluster (VSC). The authors gratefully acknowledge the Gauss Centre for Supercomputing e.V. (www.gauss-centre.eu) for funding this project by providing computing time on the GCS Supercomputer SuperMUC-NG at Leibniz Supercomputing Centre (www.lrz.de).
\end{acknowledgments}

\bibliography{Literature}

\begin{thebibliography}{50}%
\makeatletter
\providecommand \@ifxundefined [1]{%
 \@ifx{#1\undefined}
}%
\providecommand \@ifnum [1]{%
 \ifnum #1\expandafter \@firstoftwo
 \else \expandafter \@secondoftwo
 \fi
}%
\providecommand \@ifx [1]{%
 \ifx #1\expandafter \@firstoftwo
 \else \expandafter \@secondoftwo
 \fi
}%
\providecommand \natexlab [1]{#1}%
\providecommand \enquote  [1]{``#1''}%
\providecommand \bibnamefont  [1]{#1}%
\providecommand \bibfnamefont [1]{#1}%
\providecommand \citenamefont [1]{#1}%
\providecommand \href@noop [0]{\@secondoftwo}%
\providecommand \href [0]{\begingroup \@sanitize@url \@href}%
\providecommand \@href[1]{\@@startlink{#1}\@@href}%
\providecommand \@@href[1]{\endgroup#1\@@endlink}%
\providecommand \@sanitize@url [0]{\catcode `\\12\catcode `\$12\catcode
  `\&12\catcode `\#12\catcode `\^12\catcode `\_12\catcode `\%12\relax}%
\providecommand \@@startlink[1]{}%
\providecommand \@@endlink[0]{}%
\providecommand \url  [0]{\begingroup\@sanitize@url \@url }%
\providecommand \@url [1]{\endgroup\@href {#1}{\urlprefix }}%
\providecommand \urlprefix  [0]{URL }%
\providecommand \Eprint [0]{\href }%
\providecommand \doibase [0]{http://dx.doi.org/}%
\providecommand \selectlanguage [0]{\@gobble}%
\providecommand \bibinfo  [0]{\@secondoftwo}%
\providecommand \bibfield  [0]{\@secondoftwo}%
\providecommand \translation [1]{[#1]}%
\providecommand \BibitemOpen [0]{}%
\providecommand \bibitemStop [0]{}%
\providecommand \bibitemNoStop [0]{.\EOS\space}%
\providecommand \EOS [0]{\spacefactor3000\relax}%
\providecommand \BibitemShut  [1]{\csname bibitem#1\endcsname}%
\let\auto@bib@innerbib\@empty
\bibitem [{\citenamefont {Imada}\ \emph {et~al.}(1998)\citenamefont {Imada},
  \citenamefont {Fujimori},\ and\ \citenamefont {Tokura}}]{RevModPhys.70.1039}%
  \BibitemOpen
  \bibfield  {author} {\bibinfo {author} {\bibfnamefont {M.}~\bibnamefont
  {Imada}}, \bibinfo {author} {\bibfnamefont {A.}~\bibnamefont {Fujimori}}, \
  and\ \bibinfo {author} {\bibfnamefont {Y.}~\bibnamefont {Tokura}},\ }\href
  {\doibase 10.1103/RevModPhys.70.1039} {\bibfield  {journal} {\bibinfo
  {journal} {Rev. Mod. Phys.}\ }\textbf {\bibinfo {volume} {70}},\ \bibinfo
  {pages} {1039} (\bibinfo {year} {1998})}\BibitemShut {NoStop}%
\bibitem [{\citenamefont {McWhan}\ \emph {et~al.}(1973)\citenamefont {McWhan},
  \citenamefont {Menth}, \citenamefont {Remeika}, \citenamefont {Brinkman},\
  and\ \citenamefont {Rice}}]{PhysRevB.7.1920}%
  \BibitemOpen
  \bibfield  {author} {\bibinfo {author} {\bibfnamefont {D.~B.}\ \bibnamefont
  {McWhan}}, \bibinfo {author} {\bibfnamefont {A.}~\bibnamefont {Menth}},
  \bibinfo {author} {\bibfnamefont {J.~P.}\ \bibnamefont {Remeika}}, \bibinfo
  {author} {\bibfnamefont {W.~F.}\ \bibnamefont {Brinkman}}, \ and\ \bibinfo
  {author} {\bibfnamefont {T.~M.}\ \bibnamefont {Rice}},\ }\href {\doibase
  10.1103/PhysRevB.7.1920} {\bibfield  {journal} {\bibinfo  {journal} {Phys.
  Rev. B}\ }\textbf {\bibinfo {volume} {7}},\ \bibinfo {pages} {1920} (\bibinfo
  {year} {1973})}\BibitemShut {NoStop}%
\bibitem [{\citenamefont {Hansmann}\ \emph {et~al.}()\citenamefont {Hansmann},
  \citenamefont {Toschi}, \citenamefont {Sangiovanni}, \citenamefont
  {Saha‐Dasgupta}, \citenamefont {Lupi}, \citenamefont {Marsi},\ and\
  \citenamefont {Held}}]{pssb.201248476}%
  \BibitemOpen
  \bibfield  {author} {\bibinfo {author} {\bibfnamefont {P.}~\bibnamefont
  {Hansmann}}, \bibinfo {author} {\bibfnamefont {A.}~\bibnamefont {Toschi}},
  \bibinfo {author} {\bibfnamefont {G.}~\bibnamefont {Sangiovanni}}, \bibinfo
  {author} {\bibfnamefont {T.}~\bibnamefont {Saha‐Dasgupta}}, \bibinfo
  {author} {\bibfnamefont {S.}~\bibnamefont {Lupi}}, \bibinfo {author}
  {\bibfnamefont {M.}~\bibnamefont {Marsi}}, \ and\ \bibinfo {author}
  {\bibfnamefont {K.}~\bibnamefont {Held}},\ }\href {\doibase
  10.1002/pssb.201248476} {\bibfield  {journal} {\bibinfo  {journal} {physica
  status solidi (b)}\ }\textbf {\bibinfo {volume} {250}},\ \bibinfo {pages}
  {1251}}\BibitemShut {NoStop}%
\bibitem [{\citenamefont {Jackeli}\ and\ \citenamefont
  {Khaliullin}(2009)}]{PhysRevLett.102.017205}%
  \BibitemOpen
  \bibfield  {author} {\bibinfo {author} {\bibfnamefont {G.}~\bibnamefont
  {Jackeli}}\ and\ \bibinfo {author} {\bibfnamefont {G.}~\bibnamefont
  {Khaliullin}},\ }\href {\doibase 10.1103/PhysRevLett.102.017205} {\bibfield
  {journal} {\bibinfo  {journal} {Phys. Rev. Lett.}\ }\textbf {\bibinfo
  {volume} {102}},\ \bibinfo {pages} {017205} (\bibinfo {year}
  {2009})}\BibitemShut {NoStop}%
\bibitem [{\citenamefont {Martins}\ and\ \citenamefont
  {Biermann}(2017)}]{JPhysCM.29.263001}%
  \BibitemOpen
  \bibfield  {author} {\bibinfo {author} {\bibfnamefont {M.}~\bibnamefont
  {Martins}, \bibfnamefont {C.~Aichhorn}}\ and\ \bibinfo {author}
  {\bibfnamefont {S.}~\bibnamefont {Biermann}},\ }\href@noop {} {\bibfield
  {journal} {\bibinfo  {journal} {J. Phys.: Condens. Matter}\ }\textbf
  {\bibinfo {volume} {29}},\ \bibinfo {pages} {263001} (\bibinfo {year}
  {2017})}\BibitemShut {NoStop}%
\bibitem [{\citenamefont {Martins}\ \emph {et~al.}(2011)\citenamefont
  {Martins}, \citenamefont {Aichhorn}, \citenamefont {Vaugier},\ and\
  \citenamefont {Biermann}}]{PhysRevLett.107.266404}%
  \BibitemOpen
  \bibfield  {author} {\bibinfo {author} {\bibfnamefont {C.}~\bibnamefont
  {Martins}}, \bibinfo {author} {\bibfnamefont {M.}~\bibnamefont {Aichhorn}},
  \bibinfo {author} {\bibfnamefont {L.}~\bibnamefont {Vaugier}}, \ and\
  \bibinfo {author} {\bibfnamefont {S.}~\bibnamefont {Biermann}},\ }\href
  {\doibase 10.1103/PhysRevLett.107.266404} {\bibfield  {journal} {\bibinfo
  {journal} {Phys. Rev. Lett.}\ }\textbf {\bibinfo {volume} {107}},\ \bibinfo
  {pages} {266404} (\bibinfo {year} {2011})}\BibitemShut {NoStop}%
\bibitem [{\citenamefont {Kim}\ \emph {et~al.}(2008)\citenamefont {Kim},
  \citenamefont {Jin}, \citenamefont {Moon}, \citenamefont {Kim}, \citenamefont
  {Park}, \citenamefont {Leem}, \citenamefont {Yu}, \citenamefont {Noh},
  \citenamefont {Kim}, \citenamefont {Oh}, \citenamefont {Park}, \citenamefont
  {Durairaj}, \citenamefont {Cao},\ and\ \citenamefont
  {Rotenberg}}]{PhysRevLett.101.076402}%
  \BibitemOpen
  \bibfield  {author} {\bibinfo {author} {\bibfnamefont {B.~J.}\ \bibnamefont
  {Kim}}, \bibinfo {author} {\bibfnamefont {H.}~\bibnamefont {Jin}}, \bibinfo
  {author} {\bibfnamefont {S.~J.}\ \bibnamefont {Moon}}, \bibinfo {author}
  {\bibfnamefont {J.-Y.}\ \bibnamefont {Kim}}, \bibinfo {author} {\bibfnamefont
  {B.-G.}\ \bibnamefont {Park}}, \bibinfo {author} {\bibfnamefont {C.~S.}\
  \bibnamefont {Leem}}, \bibinfo {author} {\bibfnamefont {J.}~\bibnamefont
  {Yu}}, \bibinfo {author} {\bibfnamefont {T.~W.}\ \bibnamefont {Noh}},
  \bibinfo {author} {\bibfnamefont {C.}~\bibnamefont {Kim}}, \bibinfo {author}
  {\bibfnamefont {S.-J.}\ \bibnamefont {Oh}}, \bibinfo {author} {\bibfnamefont
  {J.-H.}\ \bibnamefont {Park}}, \bibinfo {author} {\bibfnamefont
  {V.}~\bibnamefont {Durairaj}}, \bibinfo {author} {\bibfnamefont
  {G.}~\bibnamefont {Cao}}, \ and\ \bibinfo {author} {\bibfnamefont
  {E.}~\bibnamefont {Rotenberg}},\ }\href {\doibase
  10.1103/PhysRevLett.101.076402} {\bibfield  {journal} {\bibinfo  {journal}
  {Phys. Rev. Lett.}\ }\textbf {\bibinfo {volume} {101}},\ \bibinfo {pages}
  {076402} (\bibinfo {year} {2008})}\BibitemShut {NoStop}%
\bibitem [{\citenamefont {Liu}\ \emph {et~al.}(2015)\citenamefont {Liu},
  \citenamefont {Khmelevskyi}, \citenamefont {Kim}, \citenamefont {Marsman},
  \citenamefont {Li}, \citenamefont {Chen}, \citenamefont {Sarma},
  \citenamefont {Kresse},\ and\ \citenamefont
  {Franchini}}]{PhysRevB.92.054428}%
  \BibitemOpen
  \bibfield  {author} {\bibinfo {author} {\bibfnamefont {P.}~\bibnamefont
  {Liu}}, \bibinfo {author} {\bibfnamefont {S.}~\bibnamefont {Khmelevskyi}},
  \bibinfo {author} {\bibfnamefont {B.}~\bibnamefont {Kim}}, \bibinfo {author}
  {\bibfnamefont {M.}~\bibnamefont {Marsman}}, \bibinfo {author} {\bibfnamefont
  {D.}~\bibnamefont {Li}}, \bibinfo {author} {\bibfnamefont {X.-Q.}\
  \bibnamefont {Chen}}, \bibinfo {author} {\bibfnamefont {D.~D.}\ \bibnamefont
  {Sarma}}, \bibinfo {author} {\bibfnamefont {G.}~\bibnamefont {Kresse}}, \
  and\ \bibinfo {author} {\bibfnamefont {C.}~\bibnamefont {Franchini}},\ }\href
  {\doibase 10.1103/PhysRevB.92.054428} {\bibfield  {journal} {\bibinfo
  {journal} {Phys. Rev. B}\ }\textbf {\bibinfo {volume} {92}},\ \bibinfo
  {pages} {054428} (\bibinfo {year} {2015})}\BibitemShut {NoStop}%
\bibitem [{\citenamefont {Kim}\ \emph {et~al.}(2017)\citenamefont {Kim},
  \citenamefont {Liu},\ and\ \citenamefont {Franchini}}]{PhysRevB.95.115111}%
  \BibitemOpen
  \bibfield  {author} {\bibinfo {author} {\bibfnamefont {B.}~\bibnamefont
  {Kim}}, \bibinfo {author} {\bibfnamefont {P.}~\bibnamefont {Liu}}, \ and\
  \bibinfo {author} {\bibfnamefont {C.}~\bibnamefont {Franchini}},\ }\href
  {\doibase 10.1103/PhysRevB.95.115111} {\bibfield  {journal} {\bibinfo
  {journal} {Phys. Rev. B}\ }\textbf {\bibinfo {volume} {95}},\ \bibinfo
  {pages} {115111} (\bibinfo {year} {2017})}\BibitemShut {NoStop}%
\bibitem [{\citenamefont {Lu}\ \emph {et~al.}(2017)\citenamefont {Lu},
  \citenamefont {Song}, \citenamefont {Liu}, \citenamefont {Reyes},
  \citenamefont {Kuhns}, \citenamefont {Lee}, \citenamefont {Fisher},\ and\
  \citenamefont {Mitrović}}]{NatComm.8.14407}%
  \BibitemOpen
  \bibfield  {author} {\bibinfo {author} {\bibfnamefont {L.}~\bibnamefont
  {Lu}}, \bibinfo {author} {\bibfnamefont {M.}~\bibnamefont {Song}}, \bibinfo
  {author} {\bibfnamefont {W.}~\bibnamefont {Liu}}, \bibinfo {author}
  {\bibfnamefont {A.~P.}\ \bibnamefont {Reyes}}, \bibinfo {author}
  {\bibfnamefont {P.}~\bibnamefont {Kuhns}}, \bibinfo {author} {\bibfnamefont
  {H.~O.}\ \bibnamefont {Lee}}, \bibinfo {author} {\bibfnamefont {I.~R.}\
  \bibnamefont {Fisher}}, \ and\ \bibinfo {author} {\bibfnamefont {V.~F.}\
  \bibnamefont {Mitrović}},\ }\href@noop {} {\bibfield  {journal} {\bibinfo
  {journal} {Nature Communications}\ }\textbf {\bibinfo {volume} {8}},\
  \bibinfo {pages} {14407} (\bibinfo {year} {2017})}\BibitemShut {NoStop}%
\bibitem [{\citenamefont {Kitagawa}\ \emph {et~al.}(2018)\citenamefont
  {Kitagawa}, \citenamefont {Takayama}, \citenamefont {Matsumoto},
  \citenamefont {Kato}, \citenamefont {Takano}, \citenamefont {Kishimoto},
  \citenamefont {Bette}, \citenamefont {Dinnebier}, \citenamefont {Jackeli},\
  and\ \citenamefont {Takagi}}]{Nature.554.341}%
  \BibitemOpen
  \bibfield  {author} {\bibinfo {author} {\bibfnamefont {K.}~\bibnamefont
  {Kitagawa}}, \bibinfo {author} {\bibfnamefont {T.}~\bibnamefont {Takayama}},
  \bibinfo {author} {\bibfnamefont {Y.}~\bibnamefont {Matsumoto}}, \bibinfo
  {author} {\bibfnamefont {A.}~\bibnamefont {Kato}}, \bibinfo {author}
  {\bibfnamefont {R.}~\bibnamefont {Takano}}, \bibinfo {author} {\bibfnamefont
  {Y.}~\bibnamefont {Kishimoto}}, \bibinfo {author} {\bibfnamefont
  {S.}~\bibnamefont {Bette}}, \bibinfo {author} {\bibfnamefont
  {R.}~\bibnamefont {Dinnebier}}, \bibinfo {author} {\bibfnamefont
  {G.}~\bibnamefont {Jackeli}}, \ and\ \bibinfo {author} {\bibfnamefont
  {H.}~\bibnamefont {Takagi}},\ }\href {https://doi.org/10.1038/nature25482}
  {\bibfield  {journal} {\bibinfo  {journal} {Nature}\ }\textbf {\bibinfo
  {volume} {554}},\ \bibinfo {pages} {341} (\bibinfo {year}
  {2018})}\BibitemShut {NoStop}%
\bibitem [{\citenamefont {Mravlje}\ \emph {et~al.}(2012)\citenamefont
  {Mravlje}, \citenamefont {Aichhorn},\ and\ \citenamefont
  {Georges}}]{PhysRevLett.108.197202}%
  \BibitemOpen
  \bibfield  {author} {\bibinfo {author} {\bibfnamefont {J.}~\bibnamefont
  {Mravlje}}, \bibinfo {author} {\bibfnamefont {M.}~\bibnamefont {Aichhorn}}, \
  and\ \bibinfo {author} {\bibfnamefont {A.}~\bibnamefont {Georges}},\ }\href
  {\doibase 10.1103/PhysRevLett.108.197202} {\bibfield  {journal} {\bibinfo
  {journal} {Phys. Rev. Lett.}\ }\textbf {\bibinfo {volume} {108}},\ \bibinfo
  {pages} {197202} (\bibinfo {year} {2012})}\BibitemShut {NoStop}%
\bibitem [{\citenamefont {de' Medici}\ \emph {et~al.}(2011)\citenamefont {de'
  Medici}, \citenamefont {Mravlje},\ and\ \citenamefont
  {Georges}}]{PhysRevLett.107.256401}%
  \BibitemOpen
  \bibfield  {author} {\bibinfo {author} {\bibfnamefont {L.}~\bibnamefont {de'
  Medici}}, \bibinfo {author} {\bibfnamefont {J.}~\bibnamefont {Mravlje}}, \
  and\ \bibinfo {author} {\bibfnamefont {A.}~\bibnamefont {Georges}},\ }\href
  {\doibase 10.1103/PhysRevLett.107.256401} {\bibfield  {journal} {\bibinfo
  {journal} {Phys. Rev. Lett.}\ }\textbf {\bibinfo {volume} {107}},\ \bibinfo
  {pages} {256401} (\bibinfo {year} {2011})}\BibitemShut {NoStop}%
\bibitem [{\citenamefont {Georges}\ \emph {et~al.}(2013)\citenamefont
  {Georges}, \citenamefont {de' Medici},\ and\ \citenamefont
  {Mravlje}}]{AnnuRevCMP83}%
  \BibitemOpen
  \bibfield  {author} {\bibinfo {author} {\bibfnamefont {A.}~\bibnamefont
  {Georges}}, \bibinfo {author} {\bibfnamefont {L.}~\bibnamefont {de' Medici}},
  \ and\ \bibinfo {author} {\bibfnamefont {J.}~\bibnamefont {Mravlje}},\ }\href
  {\doibase 10.1146/annurev-conmatphys-020911-125045} {\bibfield  {journal}
  {\bibinfo  {journal} {Annu. Rev. Condens. Matter Phys.}\ }\textbf {\bibinfo
  {volume} {4}},\ \bibinfo {pages} {137} (\bibinfo {year} {2013})}\BibitemShut
  {NoStop}%
\bibitem [{\citenamefont {Isidori}\ \emph {et~al.}(2019)\citenamefont
  {Isidori}, \citenamefont {Berovi\ifmmode~\acute{c}\else \'{c}\fi{}},
  \citenamefont {Fanfarillo}, \citenamefont {de' Medici}, \citenamefont
  {Fabrizio},\ and\ \citenamefont {Capone}}]{PhysRevLett.122.186401}%
  \BibitemOpen
  \bibfield  {author} {\bibinfo {author} {\bibfnamefont {A.}~\bibnamefont
  {Isidori}}, \bibinfo {author} {\bibfnamefont {M.}~\bibnamefont
  {Berovi\ifmmode~\acute{c}\else \'{c}\fi{}}}, \bibinfo {author} {\bibfnamefont
  {L.}~\bibnamefont {Fanfarillo}}, \bibinfo {author} {\bibfnamefont
  {L.}~\bibnamefont {de' Medici}}, \bibinfo {author} {\bibfnamefont
  {M.}~\bibnamefont {Fabrizio}}, \ and\ \bibinfo {author} {\bibfnamefont
  {M.}~\bibnamefont {Capone}},\ }\href {\doibase
  10.1103/PhysRevLett.122.186401} {\bibfield  {journal} {\bibinfo  {journal}
  {Phys. Rev. Lett.}\ }\textbf {\bibinfo {volume} {122}},\ \bibinfo {pages}
  {186401} (\bibinfo {year} {2019})}\BibitemShut {NoStop}%
\bibitem [{Note1()}]{Note1}%
  \BibitemOpen
  \bibinfo {note} {Cf. the general analysis of Refs.~\protect \rev@citealp
  {PhysRevLett.107.256401, ldmmcchapter}}\BibitemShut {NoStop}%
\bibitem [{\citenamefont {Lo~Vecchio}\ \emph {et~al.}(2013)\citenamefont
  {Lo~Vecchio}, \citenamefont {Perucchi}, \citenamefont {Di~Pietro},
  \citenamefont {Limaj}, \citenamefont {Schade}, \citenamefont {Sun},
  \citenamefont {Arai}, \citenamefont {Yamaura},\ and\ \citenamefont
  {Lupi}}]{LoVecchio2013}%
  \BibitemOpen
  \bibfield  {author} {\bibinfo {author} {\bibfnamefont {I.}~\bibnamefont
  {Lo~Vecchio}}, \bibinfo {author} {\bibfnamefont {A.}~\bibnamefont
  {Perucchi}}, \bibinfo {author} {\bibfnamefont {P.}~\bibnamefont {Di~Pietro}},
  \bibinfo {author} {\bibfnamefont {O.}~\bibnamefont {Limaj}}, \bibinfo
  {author} {\bibfnamefont {U.}~\bibnamefont {Schade}}, \bibinfo {author}
  {\bibfnamefont {Y.}~\bibnamefont {Sun}}, \bibinfo {author} {\bibfnamefont
  {M.}~\bibnamefont {Arai}}, \bibinfo {author} {\bibfnamefont {K.}~\bibnamefont
  {Yamaura}}, \ and\ \bibinfo {author} {\bibfnamefont {S.}~\bibnamefont
  {Lupi}},\ }\href {\doibase 10.1038/srep02990} {\bibfield  {journal} {\bibinfo
   {journal} {Scientific Reports}\ }\textbf {\bibinfo {volume} {3}},\ \bibinfo
  {pages} {2990} (\bibinfo {year} {2013})}\BibitemShut {NoStop}%
\bibitem [{\citenamefont {Shi}\ \emph {et~al.}(2009)\citenamefont {Shi},
  \citenamefont {Guo}, \citenamefont {Yu}, \citenamefont {Arai}, \citenamefont
  {Belik}, \citenamefont {Sato}, \citenamefont {Yamaura}, \citenamefont
  {Takayama-Muromachi}, \citenamefont {Tian}, \citenamefont {Yang},
  \citenamefont {Li}, \citenamefont {Varga}, \citenamefont {Mitchell},\ and\
  \citenamefont {Okamoto}}]{PhysRevB.80.161104}%
  \BibitemOpen
  \bibfield  {author} {\bibinfo {author} {\bibfnamefont {Y.~G.}\ \bibnamefont
  {Shi}}, \bibinfo {author} {\bibfnamefont {Y.~F.}\ \bibnamefont {Guo}},
  \bibinfo {author} {\bibfnamefont {S.}~\bibnamefont {Yu}}, \bibinfo {author}
  {\bibfnamefont {M.}~\bibnamefont {Arai}}, \bibinfo {author} {\bibfnamefont
  {A.~A.}\ \bibnamefont {Belik}}, \bibinfo {author} {\bibfnamefont
  {A.}~\bibnamefont {Sato}}, \bibinfo {author} {\bibfnamefont {K.}~\bibnamefont
  {Yamaura}}, \bibinfo {author} {\bibfnamefont {E.}~\bibnamefont
  {Takayama-Muromachi}}, \bibinfo {author} {\bibfnamefont {H.~F.}\ \bibnamefont
  {Tian}}, \bibinfo {author} {\bibfnamefont {H.~X.}\ \bibnamefont {Yang}},
  \bibinfo {author} {\bibfnamefont {J.~Q.}\ \bibnamefont {Li}}, \bibinfo
  {author} {\bibfnamefont {T.}~\bibnamefont {Varga}}, \bibinfo {author}
  {\bibfnamefont {J.~F.}\ \bibnamefont {Mitchell}}, \ and\ \bibinfo {author}
  {\bibfnamefont {S.}~\bibnamefont {Okamoto}},\ }\href {\doibase
  10.1103/PhysRevB.80.161104} {\bibfield  {journal} {\bibinfo  {journal} {Phys.
  Rev. B}\ }\textbf {\bibinfo {volume} {80}},\ \bibinfo {pages} {161104}
  (\bibinfo {year} {2009})}\BibitemShut {NoStop}%
\bibitem [{\citenamefont {Calder}\ \emph {et~al.}(2012)\citenamefont {Calder},
  \citenamefont {Garlea}, \citenamefont {McMorrow}, \citenamefont {Lumsden},
  \citenamefont {Stone}, \citenamefont {Lang}, \citenamefont {Kim},
  \citenamefont {Schlueter}, \citenamefont {Shi}, \citenamefont {Yamaura},
  \citenamefont {Sun}, \citenamefont {Tsujimoto},\ and\ \citenamefont
  {Christianson}}]{PhysRevLett.108.257209}%
  \BibitemOpen
  \bibfield  {author} {\bibinfo {author} {\bibfnamefont {S.}~\bibnamefont
  {Calder}}, \bibinfo {author} {\bibfnamefont {V.~O.}\ \bibnamefont {Garlea}},
  \bibinfo {author} {\bibfnamefont {D.~F.}\ \bibnamefont {McMorrow}}, \bibinfo
  {author} {\bibfnamefont {M.~D.}\ \bibnamefont {Lumsden}}, \bibinfo {author}
  {\bibfnamefont {M.~B.}\ \bibnamefont {Stone}}, \bibinfo {author}
  {\bibfnamefont {J.~C.}\ \bibnamefont {Lang}}, \bibinfo {author}
  {\bibfnamefont {J.-W.}\ \bibnamefont {Kim}}, \bibinfo {author} {\bibfnamefont
  {J.~A.}\ \bibnamefont {Schlueter}}, \bibinfo {author} {\bibfnamefont {Y.~G.}\
  \bibnamefont {Shi}}, \bibinfo {author} {\bibfnamefont {K.}~\bibnamefont
  {Yamaura}}, \bibinfo {author} {\bibfnamefont {Y.~S.}\ \bibnamefont {Sun}},
  \bibinfo {author} {\bibfnamefont {Y.}~\bibnamefont {Tsujimoto}}, \ and\
  \bibinfo {author} {\bibfnamefont {A.~D.}\ \bibnamefont {Christianson}},\
  }\href {\doibase 10.1103/PhysRevLett.108.257209} {\bibfield  {journal}
  {\bibinfo  {journal} {Phys. Rev. Lett.}\ }\textbf {\bibinfo {volume} {108}},\
  \bibinfo {pages} {257209} (\bibinfo {year} {2012})}\BibitemShut {NoStop}%
\bibitem [{\citenamefont {Kim}\ \emph {et~al.}(2016)\citenamefont {Kim},
  \citenamefont {Liu}, \citenamefont {Erg\"onenc}, \citenamefont {Toschi},
  \citenamefont {Khmelevskyi},\ and\ \citenamefont
  {Franchini}}]{PhysRevB.94.241113}%
  \BibitemOpen
  \bibfield  {author} {\bibinfo {author} {\bibfnamefont {B.}~\bibnamefont
  {Kim}}, \bibinfo {author} {\bibfnamefont {P.}~\bibnamefont {Liu}}, \bibinfo
  {author} {\bibfnamefont {Z.}~\bibnamefont {Erg\"onenc}}, \bibinfo {author}
  {\bibfnamefont {A.}~\bibnamefont {Toschi}}, \bibinfo {author} {\bibfnamefont
  {S.}~\bibnamefont {Khmelevskyi}}, \ and\ \bibinfo {author} {\bibfnamefont
  {C.}~\bibnamefont {Franchini}},\ }\href {\doibase 10.1103/PhysRevB.94.241113}
  {\bibfield  {journal} {\bibinfo  {journal} {Phys. Rev. B}\ }\textbf {\bibinfo
  {volume} {94}},\ \bibinfo {pages} {241113} (\bibinfo {year}
  {2016})}\BibitemShut {NoStop}%
\bibitem [{\citenamefont {Vale}\ \emph
  {et~al.}(2018{\natexlab{a}})\citenamefont {Vale}, \citenamefont {Calder},
  \citenamefont {Donnerer}, \citenamefont {Pincini}, \citenamefont {Shi},
  \citenamefont {Tsujimoto}, \citenamefont {Yamaura}, \citenamefont {Sala},
  \citenamefont {van~den Brink}, \citenamefont {Christianson},\ and\
  \citenamefont {McMorrow}}]{PhysRevLett.120.227203}%
  \BibitemOpen
  \bibfield  {author} {\bibinfo {author} {\bibfnamefont {J.~G.}\ \bibnamefont
  {Vale}}, \bibinfo {author} {\bibfnamefont {S.}~\bibnamefont {Calder}},
  \bibinfo {author} {\bibfnamefont {C.}~\bibnamefont {Donnerer}}, \bibinfo
  {author} {\bibfnamefont {D.}~\bibnamefont {Pincini}}, \bibinfo {author}
  {\bibfnamefont {Y.~G.}\ \bibnamefont {Shi}}, \bibinfo {author} {\bibfnamefont
  {Y.}~\bibnamefont {Tsujimoto}}, \bibinfo {author} {\bibfnamefont
  {K.}~\bibnamefont {Yamaura}}, \bibinfo {author} {\bibfnamefont {M.~M.}\
  \bibnamefont {Sala}}, \bibinfo {author} {\bibfnamefont {J.}~\bibnamefont
  {van~den Brink}}, \bibinfo {author} {\bibfnamefont {A.~D.}\ \bibnamefont
  {Christianson}}, \ and\ \bibinfo {author} {\bibfnamefont {D.~F.}\
  \bibnamefont {McMorrow}},\ }\href {\doibase 10.1103/PhysRevLett.120.227203}
  {\bibfield  {journal} {\bibinfo  {journal} {Phys. Rev. Lett.}\ }\textbf
  {\bibinfo {volume} {120}},\ \bibinfo {pages} {227203} (\bibinfo {year}
  {2018}{\natexlab{a}})}\BibitemShut {NoStop}%
\bibitem [{\citenamefont {Vale}\ \emph
  {et~al.}(2018{\natexlab{b}})\citenamefont {Vale}, \citenamefont {Calder},
  \citenamefont {Donnerer}, \citenamefont {Pincini}, \citenamefont {Shi},
  \citenamefont {Tsujimoto}, \citenamefont {Yamaura}, \citenamefont
  {Moretti~Sala}, \citenamefont {van~den Brink}, \citenamefont {Christianson},\
  and\ \citenamefont {McMorrow}}]{PhysRevB.97.184429}%
  \BibitemOpen
  \bibfield  {author} {\bibinfo {author} {\bibfnamefont {J.~G.}\ \bibnamefont
  {Vale}}, \bibinfo {author} {\bibfnamefont {S.}~\bibnamefont {Calder}},
  \bibinfo {author} {\bibfnamefont {C.}~\bibnamefont {Donnerer}}, \bibinfo
  {author} {\bibfnamefont {D.}~\bibnamefont {Pincini}}, \bibinfo {author}
  {\bibfnamefont {Y.~G.}\ \bibnamefont {Shi}}, \bibinfo {author} {\bibfnamefont
  {Y.}~\bibnamefont {Tsujimoto}}, \bibinfo {author} {\bibfnamefont
  {K.}~\bibnamefont {Yamaura}}, \bibinfo {author} {\bibfnamefont
  {M.}~\bibnamefont {Moretti~Sala}}, \bibinfo {author} {\bibfnamefont
  {J.}~\bibnamefont {van~den Brink}}, \bibinfo {author} {\bibfnamefont {A.~D.}\
  \bibnamefont {Christianson}}, \ and\ \bibinfo {author} {\bibfnamefont
  {D.~F.}\ \bibnamefont {McMorrow}},\ }\href {\doibase
  10.1103/PhysRevB.97.184429} {\bibfield  {journal} {\bibinfo  {journal} {Phys.
  Rev. B}\ }\textbf {\bibinfo {volume} {97}},\ \bibinfo {pages} {184429}
  (\bibinfo {year} {2018}{\natexlab{b}})}\BibitemShut {NoStop}%
\bibitem [{\citenamefont {Shi}\ \emph {et~al.}(2013)\citenamefont {Shi},
  \citenamefont {Guo}, \citenamefont {Wang}, \citenamefont {Princep},
  \citenamefont {Khalyavin}, \citenamefont {Manuel}, \citenamefont {Michiue},
  \citenamefont {Sato}, \citenamefont {Tsuda}, \citenamefont {Yu},
  \citenamefont {Arai}, \citenamefont {Shirako}, \citenamefont {Akaogi},
  \citenamefont {Wang}, \citenamefont {Yamaura},\ and\ \citenamefont
  {Boothroyd}}]{NatMat.12.1024}%
  \BibitemOpen
  \bibfield  {author} {\bibinfo {author} {\bibfnamefont {Y.}~\bibnamefont
  {Shi}}, \bibinfo {author} {\bibfnamefont {Y.}~\bibnamefont {Guo}}, \bibinfo
  {author} {\bibfnamefont {X.}~\bibnamefont {Wang}}, \bibinfo {author}
  {\bibfnamefont {A.}~\bibnamefont {Princep}}, \bibinfo {author} {\bibfnamefont
  {D.}~\bibnamefont {Khalyavin}}, \bibinfo {author} {\bibfnamefont
  {P.}~\bibnamefont {Manuel}}, \bibinfo {author} {\bibfnamefont
  {Y.}~\bibnamefont {Michiue}}, \bibinfo {author} {\bibfnamefont
  {A.}~\bibnamefont {Sato}}, \bibinfo {author} {\bibfnamefont {K.}~\bibnamefont
  {Tsuda}}, \bibinfo {author} {\bibfnamefont {S.}~\bibnamefont {Yu}}, \bibinfo
  {author} {\bibfnamefont {M.}~\bibnamefont {Arai}}, \bibinfo {author}
  {\bibfnamefont {Y.}~\bibnamefont {Shirako}}, \bibinfo {author} {\bibfnamefont
  {M.}~\bibnamefont {Akaogi}}, \bibinfo {author} {\bibfnamefont
  {N.}~\bibnamefont {Wang}}, \bibinfo {author} {\bibfnamefont {K.}~\bibnamefont
  {Yamaura}}, \ and\ \bibinfo {author} {\bibfnamefont {A.}~\bibnamefont
  {Boothroyd}},\ }\href@noop {} {\bibfield  {journal} {\bibinfo  {journal}
  {Nature Materials}\ }\textbf {\bibinfo {volume} {12}},\ \bibinfo {pages}
  {1024} (\bibinfo {year} {2013})}\BibitemShut {NoStop}%
\bibitem [{\citenamefont {Lo~Vecchio}\ \emph {et~al.}(2016)\citenamefont
  {Lo~Vecchio}, \citenamefont {Giovannetti}, \citenamefont {Autore},
  \citenamefont {Di~Pietro}, \citenamefont {Perucchi}, \citenamefont {He},
  \citenamefont {Yamaura}, \citenamefont {Capone},\ and\ \citenamefont
  {Lupi}}]{PhysRevB.93.161113}%
  \BibitemOpen
  \bibfield  {author} {\bibinfo {author} {\bibfnamefont {I.}~\bibnamefont
  {Lo~Vecchio}}, \bibinfo {author} {\bibfnamefont {G.}~\bibnamefont
  {Giovannetti}}, \bibinfo {author} {\bibfnamefont {M.}~\bibnamefont {Autore}},
  \bibinfo {author} {\bibfnamefont {P.}~\bibnamefont {Di~Pietro}}, \bibinfo
  {author} {\bibfnamefont {A.}~\bibnamefont {Perucchi}}, \bibinfo {author}
  {\bibfnamefont {J.}~\bibnamefont {He}}, \bibinfo {author} {\bibfnamefont
  {K.}~\bibnamefont {Yamaura}}, \bibinfo {author} {\bibfnamefont
  {M.}~\bibnamefont {Capone}}, \ and\ \bibinfo {author} {\bibfnamefont
  {S.}~\bibnamefont {Lupi}},\ }\href {\doibase 10.1103/PhysRevB.93.161113}
  {\bibfield  {journal} {\bibinfo  {journal} {Phys. Rev. B}\ }\textbf {\bibinfo
  {volume} {93}},\ \bibinfo {pages} {161113} (\bibinfo {year}
  {2016})}\BibitemShut {NoStop}%
\bibitem [{\citenamefont {Baldassarre}\ \emph {et~al.}(2008)\citenamefont
  {Baldassarre}, \citenamefont {Perucchi}, \citenamefont {Nicoletti},
  \citenamefont {Toschi}, \citenamefont {Sangiovanni}, \citenamefont {Held},
  \citenamefont {Capone}, \citenamefont {Ortolani}, \citenamefont {Malavasi},
  \citenamefont {Marsi}, \citenamefont {Metcalf}, \citenamefont {Postorino},\
  and\ \citenamefont {Lupi}}]{PhysRevB.77.113107}%
  \BibitemOpen
  \bibfield  {author} {\bibinfo {author} {\bibfnamefont {L.}~\bibnamefont
  {Baldassarre}}, \bibinfo {author} {\bibfnamefont {A.}~\bibnamefont
  {Perucchi}}, \bibinfo {author} {\bibfnamefont {D.}~\bibnamefont {Nicoletti}},
  \bibinfo {author} {\bibfnamefont {A.}~\bibnamefont {Toschi}}, \bibinfo
  {author} {\bibfnamefont {G.}~\bibnamefont {Sangiovanni}}, \bibinfo {author}
  {\bibfnamefont {K.}~\bibnamefont {Held}}, \bibinfo {author} {\bibfnamefont
  {M.}~\bibnamefont {Capone}}, \bibinfo {author} {\bibfnamefont
  {M.}~\bibnamefont {Ortolani}}, \bibinfo {author} {\bibfnamefont
  {L.}~\bibnamefont {Malavasi}}, \bibinfo {author} {\bibfnamefont
  {M.}~\bibnamefont {Marsi}}, \bibinfo {author} {\bibfnamefont
  {P.}~\bibnamefont {Metcalf}}, \bibinfo {author} {\bibfnamefont
  {P.}~\bibnamefont {Postorino}}, \ and\ \bibinfo {author} {\bibfnamefont
  {S.}~\bibnamefont {Lupi}},\ }\href {\doibase 10.1103/PhysRevB.77.113107}
  {\bibfield  {journal} {\bibinfo  {journal} {Phys. Rev. B}\ }\textbf {\bibinfo
  {volume} {77}},\ \bibinfo {pages} {113107} (\bibinfo {year}
  {2008})}\BibitemShut {NoStop}%
\bibitem [{Sup()}]{Suppl}%
  \BibitemOpen
  \href@noop {} {\bibinfo  {journal} {for further details, see Supplemental
  Material, which also includes Refs. \cite{PhysRevLett.77.3865, Kaltak2015,
  PhysRevB.65.035109, RevModPhys.83.349}}\ }\BibitemShut {NoStop}%
\bibitem [{\citenamefont {Kresse}\ and\ \citenamefont
  {Hafner}(1993)}]{Kresse1993}%
  \BibitemOpen
\bibfield  {journal} {  }\bibfield  {author} {\bibinfo {author} {\bibfnamefont
  {G.}~\bibnamefont {Kresse}}\ and\ \bibinfo {author} {\bibfnamefont
  {J.}~\bibnamefont {Hafner}},\ }\href {\doibase 10.1103/PhysRevB.47.558}
  {\bibfield  {journal} {\bibinfo  {journal} {Phys. Rev. B}\ }\textbf {\bibinfo
  {volume} {47}},\ \bibinfo {pages} {558} (\bibinfo {year} {1993})}\BibitemShut
  {NoStop}%
\bibitem [{\citenamefont {Kresse}\ and\ \citenamefont
  {Furthm\"uller}(1996)}]{Kresse1996}%
  \BibitemOpen
  \bibfield  {author} {\bibinfo {author} {\bibfnamefont {G.}~\bibnamefont
  {Kresse}}\ and\ \bibinfo {author} {\bibfnamefont {J.}~\bibnamefont
  {Furthm\"uller}},\ }\href {\doibase 10.1103/PhysRevB.54.11169} {\bibfield
  {journal} {\bibinfo  {journal} {Phys. Rev. B}\ }\textbf {\bibinfo {volume}
  {54}},\ \bibinfo {pages} {11169} (\bibinfo {year} {1996})}\BibitemShut
  {NoStop}%
\bibitem [{\citenamefont {Georges}\ \emph {et~al.}(1996)\citenamefont
  {Georges}, \citenamefont {Kotliar}, \citenamefont {Krauth},\ and\
  \citenamefont {Rozenberg}}]{RevModPhys.68.13}%
  \BibitemOpen
  \bibfield  {author} {\bibinfo {author} {\bibfnamefont {A.}~\bibnamefont
  {Georges}}, \bibinfo {author} {\bibfnamefont {G.}~\bibnamefont {Kotliar}},
  \bibinfo {author} {\bibfnamefont {W.}~\bibnamefont {Krauth}}, \ and\ \bibinfo
  {author} {\bibfnamefont {M.~J.}\ \bibnamefont {Rozenberg}},\ }\href {\doibase
  10.1103/RevModPhys.68.13} {\bibfield  {journal} {\bibinfo  {journal} {Rev.
  Mod. Phys.}\ }\textbf {\bibinfo {volume} {68}},\ \bibinfo {pages} {13}
  (\bibinfo {year} {1996})}\BibitemShut {NoStop}%
\bibitem [{Note2()}]{Note2}%
  \BibitemOpen
  \bibinfo {note} {In our maximally localized Wannier projection we have
  effectively downfolded both the high-lying $e_g$ orbitals of Os, which are
  located approximately $2$ eV, as well as the low-lying $2p$ orbitals of the
  O~\cite {Suppl}}\BibitemShut {NoStop}%
\bibitem [{\citenamefont {Liu}\ \emph {et~al.}(2020)\citenamefont {Liu},
  \citenamefont {He}, \citenamefont {Kim}, \citenamefont {Khmelevskyi},
  \citenamefont {Toschi}, \citenamefont {Kresse},\ and\ \citenamefont
  {Franchini}}]{Peitao2020}%
  \BibitemOpen
  \bibfield  {author} {\bibinfo {author} {\bibfnamefont {P.}~\bibnamefont
  {Liu}}, \bibinfo {author} {\bibfnamefont {J.}~\bibnamefont {He}}, \bibinfo
  {author} {\bibfnamefont {B.}~\bibnamefont {Kim}}, \bibinfo {author}
  {\bibfnamefont {S.}~\bibnamefont {Khmelevskyi}}, \bibinfo {author}
  {\bibfnamefont {A.}~\bibnamefont {Toschi}}, \bibinfo {author} {\bibfnamefont
  {G.}~\bibnamefont {Kresse}}, \ and\ \bibinfo {author} {\bibfnamefont
  {C.}~\bibnamefont {Franchini}},\ }\href {\doibase
  10.1103/PhysRevMaterials.4.045001} {\bibfield  {journal} {\bibinfo  {journal}
  {Phys. Rev. Materials}\ }\textbf {\bibinfo {volume} {4}},\ \bibinfo {pages}
  {045001} (\bibinfo {year} {2020})}\BibitemShut {NoStop}%
\bibitem [{\citenamefont {Calder}\ \emph {et~al.}(2017)\citenamefont {Calder},
  \citenamefont {Vale}, \citenamefont {Bogdanov}, \citenamefont {Donnerer},
  \citenamefont {Pincini}, \citenamefont {Moretti~Sala}, \citenamefont {Liu},
  \citenamefont {Upton}, \citenamefont {Casa}, \citenamefont {Shi},
  \citenamefont {Tsujimoto}, \citenamefont {Yamaura}, \citenamefont {Hill},
  \citenamefont {van~den Brink}, \citenamefont {McMorrow},\ and\ \citenamefont
  {Christianson}}]{PhysRevB.95.020413}%
  \BibitemOpen
  \bibfield  {author} {\bibinfo {author} {\bibfnamefont {S.}~\bibnamefont
  {Calder}}, \bibinfo {author} {\bibfnamefont {J.~G.}\ \bibnamefont {Vale}},
  \bibinfo {author} {\bibfnamefont {N.}~\bibnamefont {Bogdanov}}, \bibinfo
  {author} {\bibfnamefont {C.}~\bibnamefont {Donnerer}}, \bibinfo {author}
  {\bibfnamefont {D.}~\bibnamefont {Pincini}}, \bibinfo {author} {\bibfnamefont
  {M.}~\bibnamefont {Moretti~Sala}}, \bibinfo {author} {\bibfnamefont
  {X.}~\bibnamefont {Liu}}, \bibinfo {author} {\bibfnamefont {M.~H.}\
  \bibnamefont {Upton}}, \bibinfo {author} {\bibfnamefont {D.}~\bibnamefont
  {Casa}}, \bibinfo {author} {\bibfnamefont {Y.~G.}\ \bibnamefont {Shi}},
  \bibinfo {author} {\bibfnamefont {Y.}~\bibnamefont {Tsujimoto}}, \bibinfo
  {author} {\bibfnamefont {K.}~\bibnamefont {Yamaura}}, \bibinfo {author}
  {\bibfnamefont {J.~P.}\ \bibnamefont {Hill}}, \bibinfo {author}
  {\bibfnamefont {J.}~\bibnamefont {van~den Brink}}, \bibinfo {author}
  {\bibfnamefont {D.~F.}\ \bibnamefont {McMorrow}}, \ and\ \bibinfo {author}
  {\bibfnamefont {A.~D.}\ \bibnamefont {Christianson}},\ }\href {\doibase
  10.1103/PhysRevB.95.020413} {\bibfield  {journal} {\bibinfo  {journal} {Phys.
  Rev. B}\ }\textbf {\bibinfo {volume} {95}},\ \bibinfo {pages} {020413}
  (\bibinfo {year} {2017})}\BibitemShut {NoStop}%
\bibitem [{\citenamefont {Calder}\ \emph {et~al.}(2016)\citenamefont {Calder},
  \citenamefont {Vale}, \citenamefont {Bogdanov}, \citenamefont {Liu},
  \citenamefont {Donnerer}, \citenamefont {Upton}, \citenamefont {Casa},
  \citenamefont {Said}, \citenamefont {Lumsden}, \citenamefont {Zhao},
  \citenamefont {Yan}, \citenamefont {Mandrus}, \citenamefont {Nishimoto},
  \citenamefont {van~den Brink}, \citenamefont {Hill}, \citenamefont
  {McMorrow},\ and\ \citenamefont {Christianson}}]{Calder2016}%
  \BibitemOpen
  \bibfield  {author} {\bibinfo {author} {\bibfnamefont {S.}~\bibnamefont
  {Calder}}, \bibinfo {author} {\bibfnamefont {J.~G.}\ \bibnamefont {Vale}},
  \bibinfo {author} {\bibfnamefont {N.~A.}\ \bibnamefont {Bogdanov}}, \bibinfo
  {author} {\bibfnamefont {X.}~\bibnamefont {Liu}}, \bibinfo {author}
  {\bibfnamefont {C.}~\bibnamefont {Donnerer}}, \bibinfo {author}
  {\bibfnamefont {M.~H.}\ \bibnamefont {Upton}}, \bibinfo {author}
  {\bibfnamefont {D.}~\bibnamefont {Casa}}, \bibinfo {author} {\bibfnamefont
  {A.~H.}\ \bibnamefont {Said}}, \bibinfo {author} {\bibfnamefont {M.~D.}\
  \bibnamefont {Lumsden}}, \bibinfo {author} {\bibfnamefont {Z.}~\bibnamefont
  {Zhao}}, \bibinfo {author} {\bibfnamefont {J.-Q.}\ \bibnamefont {Yan}},
  \bibinfo {author} {\bibfnamefont {D.}~\bibnamefont {Mandrus}}, \bibinfo
  {author} {\bibfnamefont {S.}~\bibnamefont {Nishimoto}}, \bibinfo {author}
  {\bibfnamefont {J.}~\bibnamefont {van~den Brink}}, \bibinfo {author}
  {\bibfnamefont {J.~P.}\ \bibnamefont {Hill}}, \bibinfo {author}
  {\bibfnamefont {D.~F.}\ \bibnamefont {McMorrow}}, \ and\ \bibinfo {author}
  {\bibfnamefont {A.~D.}\ \bibnamefont {Christianson}},\ }\href {\doibase
  10.1038/ncomms11651} {\bibfield  {journal} {\bibinfo  {journal} {Nature
  Communications}\ }\textbf {\bibinfo {volume} {7}},\ \bibinfo {pages} {11651}
  (\bibinfo {year} {2016})}\BibitemShut {NoStop}%
\bibitem [{\citenamefont {Taylor}\ \emph {et~al.}(2016)\citenamefont {Taylor},
  \citenamefont {Morrow}, \citenamefont {Fishman}, \citenamefont {Calder},
  \citenamefont {Kolesnikov}, \citenamefont {Lumsden}, \citenamefont
  {Woodward},\ and\ \citenamefont {Christianson}}]{PhysRevB.93.220408}%
  \BibitemOpen
  \bibfield  {author} {\bibinfo {author} {\bibfnamefont {A.~E.}\ \bibnamefont
  {Taylor}}, \bibinfo {author} {\bibfnamefont {R.}~\bibnamefont {Morrow}},
  \bibinfo {author} {\bibfnamefont {R.~S.}\ \bibnamefont {Fishman}}, \bibinfo
  {author} {\bibfnamefont {S.}~\bibnamefont {Calder}}, \bibinfo {author}
  {\bibfnamefont {A.~I.}\ \bibnamefont {Kolesnikov}}, \bibinfo {author}
  {\bibfnamefont {M.~D.}\ \bibnamefont {Lumsden}}, \bibinfo {author}
  {\bibfnamefont {P.~M.}\ \bibnamefont {Woodward}}, \ and\ \bibinfo {author}
  {\bibfnamefont {A.~D.}\ \bibnamefont {Christianson}},\ }\href {\doibase
  10.1103/PhysRevB.93.220408} {\bibfield  {journal} {\bibinfo  {journal} {Phys.
  Rev. B}\ }\textbf {\bibinfo {volume} {93}},\ \bibinfo {pages} {220408}
  (\bibinfo {year} {2016})}\BibitemShut {NoStop}%
\bibitem [{\citenamefont {Kermarrec}\ \emph {et~al.}(2015)\citenamefont
  {Kermarrec}, \citenamefont {Marjerrison}, \citenamefont {Thompson},
  \citenamefont {Maharaj}, \citenamefont {Levin}, \citenamefont {Kroeker},
  \citenamefont {Granroth}, \citenamefont {Flacau}, \citenamefont {Yamani},
  \citenamefont {Greedan},\ and\ \citenamefont {Gaulin}}]{PhysRevB.91.075133}%
  \BibitemOpen
  \bibfield  {author} {\bibinfo {author} {\bibfnamefont {E.}~\bibnamefont
  {Kermarrec}}, \bibinfo {author} {\bibfnamefont {C.~A.}\ \bibnamefont
  {Marjerrison}}, \bibinfo {author} {\bibfnamefont {C.~M.}\ \bibnamefont
  {Thompson}}, \bibinfo {author} {\bibfnamefont {D.~D.}\ \bibnamefont
  {Maharaj}}, \bibinfo {author} {\bibfnamefont {K.}~\bibnamefont {Levin}},
  \bibinfo {author} {\bibfnamefont {S.}~\bibnamefont {Kroeker}}, \bibinfo
  {author} {\bibfnamefont {G.~E.}\ \bibnamefont {Granroth}}, \bibinfo {author}
  {\bibfnamefont {R.}~\bibnamefont {Flacau}}, \bibinfo {author} {\bibfnamefont
  {Z.}~\bibnamefont {Yamani}}, \bibinfo {author} {\bibfnamefont {J.~E.}\
  \bibnamefont {Greedan}}, \ and\ \bibinfo {author} {\bibfnamefont {B.~D.}\
  \bibnamefont {Gaulin}},\ }\href {\doibase 10.1103/PhysRevB.91.075133}
  {\bibfield  {journal} {\bibinfo  {journal} {Phys. Rev. B}\ }\textbf {\bibinfo
  {volume} {91}},\ \bibinfo {pages} {075133} (\bibinfo {year}
  {2015})}\BibitemShut {NoStop}%
\bibitem [{\citenamefont {Wallerberger}\ \emph {et~al.}(2019)\citenamefont
  {Wallerberger}, \citenamefont {Hausoel}, \citenamefont {Gunacker},
  \citenamefont {Kowalski}, \citenamefont {Parragh}, \citenamefont {Goth},
  \citenamefont {Held},\ and\ \citenamefont
  {Sangiovanni}}]{Comput.Phys.Commun.235.388}%
  \BibitemOpen
  \bibfield  {author} {\bibinfo {author} {\bibfnamefont {M.}~\bibnamefont
  {Wallerberger}}, \bibinfo {author} {\bibfnamefont {A.}~\bibnamefont
  {Hausoel}}, \bibinfo {author} {\bibfnamefont {P.}~\bibnamefont {Gunacker}},
  \bibinfo {author} {\bibfnamefont {A.}~\bibnamefont {Kowalski}}, \bibinfo
  {author} {\bibfnamefont {N.}~\bibnamefont {Parragh}}, \bibinfo {author}
  {\bibfnamefont {F.}~\bibnamefont {Goth}}, \bibinfo {author} {\bibfnamefont
  {K.}~\bibnamefont {Held}}, \ and\ \bibinfo {author} {\bibfnamefont
  {G.}~\bibnamefont {Sangiovanni}},\ }\href@noop {} {\bibfield  {journal}
  {\bibinfo  {journal} {Comput. Phys. Commun.}\ }\textbf {\bibinfo {volume}
  {235}},\ \bibinfo {pages} {388} (\bibinfo {year} {2019})}\BibitemShut
  {NoStop}%
\bibitem [{\citenamefont {Honerkamp}(2018)}]{PhysRevB.98.155132}%
  \BibitemOpen
  \bibfield  {author} {\bibinfo {author} {\bibfnamefont {C.}~\bibnamefont
  {Honerkamp}},\ }\href {\doibase 10.1103/PhysRevB.98.155132} {\bibfield
  {journal} {\bibinfo  {journal} {Phys. Rev. B}\ }\textbf {\bibinfo {volume}
  {98}},\ \bibinfo {pages} {155132} (\bibinfo {year} {2018})}\BibitemShut
  {NoStop}%
\bibitem [{\citenamefont {Honerkamp}\ \emph {et~al.}(2018)\citenamefont
  {Honerkamp}, \citenamefont {Shinaoka}, \citenamefont {Assaad},\ and\
  \citenamefont {Werner}}]{PhysRevB.98.235151}%
  \BibitemOpen
  \bibfield  {author} {\bibinfo {author} {\bibfnamefont {C.}~\bibnamefont
  {Honerkamp}}, \bibinfo {author} {\bibfnamefont {H.}~\bibnamefont {Shinaoka}},
  \bibinfo {author} {\bibfnamefont {F.~F.}\ \bibnamefont {Assaad}}, \ and\
  \bibinfo {author} {\bibfnamefont {P.}~\bibnamefont {Werner}},\ }\href
  {\doibase 10.1103/PhysRevB.98.235151} {\bibfield  {journal} {\bibinfo
  {journal} {Phys. Rev. B}\ }\textbf {\bibinfo {volume} {98}},\ \bibinfo
  {pages} {235151} (\bibinfo {year} {2018})}\BibitemShut {NoStop}%
\bibitem [{\citenamefont {Levy}\ \emph {et~al.}(2017)\citenamefont {Levy},
  \citenamefont {LeBlanc},\ and\ \citenamefont {Gull}}]{LEVY2017149}%
  \BibitemOpen
  \bibfield  {author} {\bibinfo {author} {\bibfnamefont {R.}~\bibnamefont
  {Levy}}, \bibinfo {author} {\bibfnamefont {J.}~\bibnamefont {LeBlanc}}, \
  and\ \bibinfo {author} {\bibfnamefont {E.}~\bibnamefont {Gull}},\ }\href
  {\doibase https://doi.org/10.1016/j.cpc.2017.01.018} {\bibfield  {journal}
  {\bibinfo  {journal} {Computer Physics Communications}\ }\textbf {\bibinfo
  {volume} {215}},\ \bibinfo {pages} {149 } (\bibinfo {year}
  {2017})}\BibitemShut {NoStop}%
\bibitem [{\citenamefont {Gaenko}\ \emph {et~al.}(2017)\citenamefont {Gaenko},
  \citenamefont {Antipov}, \citenamefont {Carcassi}, \citenamefont {Chen},
  \citenamefont {Chen}, \citenamefont {Dong}, \citenamefont {Gamper},
  \citenamefont {Gukelberger}, \citenamefont {Igarashi}, \citenamefont
  {Iskakov}, \citenamefont {Könz}, \citenamefont {LeBlanc}, \citenamefont
  {Levy}, \citenamefont {Ma}, \citenamefont {Paki}, \citenamefont {Shinaoka},
  \citenamefont {Todo}, \citenamefont {Troyer},\ and\ \citenamefont
  {Gull}}]{GAENKO2017235}%
  \BibitemOpen
  \bibfield  {author} {\bibinfo {author} {\bibfnamefont {A.}~\bibnamefont
  {Gaenko}}, \bibinfo {author} {\bibfnamefont {A.}~\bibnamefont {Antipov}},
  \bibinfo {author} {\bibfnamefont {G.}~\bibnamefont {Carcassi}}, \bibinfo
  {author} {\bibfnamefont {T.}~\bibnamefont {Chen}}, \bibinfo {author}
  {\bibfnamefont {X.}~\bibnamefont {Chen}}, \bibinfo {author} {\bibfnamefont
  {Q.}~\bibnamefont {Dong}}, \bibinfo {author} {\bibfnamefont {L.}~\bibnamefont
  {Gamper}}, \bibinfo {author} {\bibfnamefont {J.}~\bibnamefont {Gukelberger}},
  \bibinfo {author} {\bibfnamefont {R.}~\bibnamefont {Igarashi}}, \bibinfo
  {author} {\bibfnamefont {S.}~\bibnamefont {Iskakov}}, \bibinfo {author}
  {\bibfnamefont {M.}~\bibnamefont {Könz}}, \bibinfo {author} {\bibfnamefont
  {J.}~\bibnamefont {LeBlanc}}, \bibinfo {author} {\bibfnamefont
  {R.}~\bibnamefont {Levy}}, \bibinfo {author} {\bibfnamefont {P.}~\bibnamefont
  {Ma}}, \bibinfo {author} {\bibfnamefont {J.}~\bibnamefont {Paki}}, \bibinfo
  {author} {\bibfnamefont {H.}~\bibnamefont {Shinaoka}}, \bibinfo {author}
  {\bibfnamefont {S.}~\bibnamefont {Todo}}, \bibinfo {author} {\bibfnamefont
  {M.}~\bibnamefont {Troyer}}, \ and\ \bibinfo {author} {\bibfnamefont
  {E.}~\bibnamefont {Gull}},\ }\href {\doibase
  https://doi.org/10.1016/j.cpc.2016.12.009} {\bibfield  {journal} {\bibinfo
  {journal} {Computer Physics Communications}\ }\textbf {\bibinfo {volume}
  {213}},\ \bibinfo {pages} {235 } (\bibinfo {year} {2017})}\BibitemShut
  {NoStop}%
\bibitem [{\citenamefont {Bauer}\ \emph {et~al.}(2011)\citenamefont {Bauer},
  \citenamefont {Carr}, \citenamefont {Evertz}, \citenamefont {Feiguin},
  \citenamefont {Freire}, \citenamefont {Fuchs}, \citenamefont {Gamper},
  \citenamefont {Gukelberger}, \citenamefont {Gull}, \citenamefont {Guertler},
  \citenamefont {Hehn}, \citenamefont {Igarashi}, \citenamefont {Isakov},
  \citenamefont {Koop}, \citenamefont {Ma}, \citenamefont {Mates},
  \citenamefont {Matsuo}, \citenamefont {Parcollet}, \citenamefont
  {Pawłowski}, \citenamefont {Picon}, \citenamefont {Pollet}, \citenamefont
  {Santos}, \citenamefont {Scarola}, \citenamefont {Schollwöck}, \citenamefont
  {Silva}, \citenamefont {Surer}, \citenamefont {Todo}, \citenamefont {Trebst},
  \citenamefont {Troyer}, \citenamefont {Wall}, \citenamefont {Werner},\ and\
  \citenamefont {Wessel}}]{1742-5468-2011-05-P05001}%
  \BibitemOpen
  \bibfield  {author} {\bibinfo {author} {\bibfnamefont {B.}~\bibnamefont
  {Bauer}}, \bibinfo {author} {\bibfnamefont {L.~D.}\ \bibnamefont {Carr}},
  \bibinfo {author} {\bibfnamefont {H.~G.}\ \bibnamefont {Evertz}}, \bibinfo
  {author} {\bibfnamefont {A.}~\bibnamefont {Feiguin}}, \bibinfo {author}
  {\bibfnamefont {J.}~\bibnamefont {Freire}}, \bibinfo {author} {\bibfnamefont
  {S.}~\bibnamefont {Fuchs}}, \bibinfo {author} {\bibfnamefont
  {L.}~\bibnamefont {Gamper}}, \bibinfo {author} {\bibfnamefont
  {J.}~\bibnamefont {Gukelberger}}, \bibinfo {author} {\bibfnamefont
  {E.}~\bibnamefont {Gull}}, \bibinfo {author} {\bibfnamefont {S.}~\bibnamefont
  {Guertler}}, \bibinfo {author} {\bibfnamefont {A.}~\bibnamefont {Hehn}},
  \bibinfo {author} {\bibfnamefont {R.}~\bibnamefont {Igarashi}}, \bibinfo
  {author} {\bibfnamefont {S.~V.}\ \bibnamefont {Isakov}}, \bibinfo {author}
  {\bibfnamefont {D.}~\bibnamefont {Koop}}, \bibinfo {author} {\bibfnamefont
  {P.~N.}\ \bibnamefont {Ma}}, \bibinfo {author} {\bibfnamefont
  {P.}~\bibnamefont {Mates}}, \bibinfo {author} {\bibfnamefont
  {H.}~\bibnamefont {Matsuo}}, \bibinfo {author} {\bibfnamefont
  {O.}~\bibnamefont {Parcollet}}, \bibinfo {author} {\bibfnamefont
  {G.}~\bibnamefont {Pawłowski}}, \bibinfo {author} {\bibfnamefont {J.~D.}\
  \bibnamefont {Picon}}, \bibinfo {author} {\bibfnamefont {L.}~\bibnamefont
  {Pollet}}, \bibinfo {author} {\bibfnamefont {E.}~\bibnamefont {Santos}},
  \bibinfo {author} {\bibfnamefont {V.~W.}\ \bibnamefont {Scarola}}, \bibinfo
  {author} {\bibfnamefont {U.}~\bibnamefont {Schollwöck}}, \bibinfo {author}
  {\bibfnamefont {C.}~\bibnamefont {Silva}}, \bibinfo {author} {\bibfnamefont
  {B.}~\bibnamefont {Surer}}, \bibinfo {author} {\bibfnamefont
  {S.}~\bibnamefont {Todo}}, \bibinfo {author} {\bibfnamefont {S.}~\bibnamefont
  {Trebst}}, \bibinfo {author} {\bibfnamefont {M.}~\bibnamefont {Troyer}},
  \bibinfo {author} {\bibfnamefont {M.~L.}\ \bibnamefont {Wall}}, \bibinfo
  {author} {\bibfnamefont {P.}~\bibnamefont {Werner}}, \ and\ \bibinfo {author}
  {\bibfnamefont {S.}~\bibnamefont {Wessel}},\ }\href
  {http://stacks.iop.org/1742-5468/2011/i=05/a=P05001} {\bibfield  {journal}
  {\bibinfo  {journal} {Journal of Statistical Mechanics: Theory and
  Experiment}\ }\textbf {\bibinfo {volume} {2011}},\ \bibinfo {pages} {P05001}
  (\bibinfo {year} {2011})}\BibitemShut {NoStop}%
\bibitem [{Note3()}]{Note3}%
  \BibitemOpen
  \bibinfo {note} {At the other temperatures shown n Fig.~\ref
  {Fig:FullCoulomb} we did not observe a coexistence of insulating and metallic
  DMFT solutions for LiOsO$_3$.}\BibitemShut {Stop}%
\bibitem [{\citenamefont {Horvat}\ \emph {et~al.}(2017)\citenamefont {Horvat},
  \citenamefont {Pourovskii}, \citenamefont {Aichhorn},\ and\ \citenamefont
  {Mravlje}}]{PhysRevB.95.205115}%
  \BibitemOpen
  \bibfield  {author} {\bibinfo {author} {\bibfnamefont {A.}~\bibnamefont
  {Horvat}}, \bibinfo {author} {\bibfnamefont {L.}~\bibnamefont {Pourovskii}},
  \bibinfo {author} {\bibfnamefont {M.}~\bibnamefont {Aichhorn}}, \ and\
  \bibinfo {author} {\bibfnamefont {J.}~\bibnamefont {Mravlje}},\ }\href
  {\doibase 10.1103/PhysRevB.95.205115} {\bibfield  {journal} {\bibinfo
  {journal} {Phys. Rev. B}\ }\textbf {\bibinfo {volume} {95}},\ \bibinfo
  {pages} {205115} (\bibinfo {year} {2017})}\BibitemShut {NoStop}%
\bibitem [{\citenamefont {Takeda}\ and\ \citenamefont
  {Ohara}(1974)}]{JPSJ.37.275}%
  \BibitemOpen
  \bibfield  {author} {\bibinfo {author} {\bibfnamefont {T.}~\bibnamefont
  {Takeda}}\ and\ \bibinfo {author} {\bibfnamefont {S.}~\bibnamefont {Ohara}},\
  }\href {\doibase 10.1143/JPSJ.37.275} {\bibfield  {journal} {\bibinfo
  {journal} {Journal of the Physical Society of Japan}\ }\textbf {\bibinfo
  {volume} {37}},\ \bibinfo {pages} {275} (\bibinfo {year} {1974})},\ \Eprint
  {http://arxiv.org/abs/https://doi.org/10.1143/JPSJ.37.275}
  {https://doi.org/10.1143/JPSJ.37.275} \BibitemShut {NoStop}%
\bibitem [{\citenamefont {S\o{}nden\aa{}}\ \emph {et~al.}(2006)\citenamefont
  {S\o{}nden\aa{}}, \citenamefont {Ravindran}, \citenamefont {St\o{}len},
  \citenamefont {Grande},\ and\ \citenamefont {Hanfland}}]{PhysRevB.74.144102}%
  \BibitemOpen
  \bibfield  {author} {\bibinfo {author} {\bibfnamefont {R.}~\bibnamefont
  {S\o{}nden\aa{}}}, \bibinfo {author} {\bibfnamefont {P.}~\bibnamefont
  {Ravindran}}, \bibinfo {author} {\bibfnamefont {S.}~\bibnamefont
  {St\o{}len}}, \bibinfo {author} {\bibfnamefont {T.}~\bibnamefont {Grande}}, \
  and\ \bibinfo {author} {\bibfnamefont {M.}~\bibnamefont {Hanfland}},\ }\href
  {\doibase 10.1103/PhysRevB.74.144102} {\bibfield  {journal} {\bibinfo
  {journal} {Phys. Rev. B}\ }\textbf {\bibinfo {volume} {74}},\ \bibinfo
  {pages} {144102} (\bibinfo {year} {2006})}\BibitemShut {NoStop}%
\bibitem [{\citenamefont {de' Medici}\ and\ \citenamefont
  {Capone}(2017)}]{ldmmcchapter}%
  \BibitemOpen
  \bibfield  {author} {\bibinfo {author} {\bibfnamefont {L.}~\bibnamefont {de'
  Medici}}\ and\ \bibinfo {author} {\bibfnamefont {M.}~\bibnamefont {Capone}},\
  }\enquote {\bibinfo {title} {Modeling many-body physics with slave-spin
  mean-field: Mott and hund's physics in fe-superconductors},}\ in\ \href
  {\doibase 10.1007/978-3-319-56117-2_4} {\emph {\bibinfo {booktitle} {The Iron
  Pnictide Superconductors: An Introduction and Overview}}},\ \bibinfo {editor}
  {edited by\ \bibinfo {editor} {\bibfnamefont {F.}~\bibnamefont {Mancini}}\
  and\ \bibinfo {editor} {\bibfnamefont {R.}~\bibnamefont {Citro}}}\ (\bibinfo
  {publisher} {Springer International Publishing},\ \bibinfo {address} {Cham},\
  \bibinfo {year} {2017})\ pp.\ \bibinfo {pages} {115--185}\BibitemShut
  {NoStop}%
\bibitem [{\citenamefont {Perdew}\ \emph {et~al.}(1996)\citenamefont {Perdew},
  \citenamefont {Burke},\ and\ \citenamefont
  {Ernzerhof}}]{PhysRevLett.77.3865}%
  \BibitemOpen
  \bibfield  {author} {\bibinfo {author} {\bibfnamefont {J.~P.}\ \bibnamefont
  {Perdew}}, \bibinfo {author} {\bibfnamefont {K.}~\bibnamefont {Burke}}, \
  and\ \bibinfo {author} {\bibfnamefont {M.}~\bibnamefont {Ernzerhof}},\ }\href
  {\doibase 10.1103/PhysRevLett.77.3865} {\bibfield  {journal} {\bibinfo
  {journal} {Phys. Rev. Lett.}\ }\textbf {\bibinfo {volume} {77}},\ \bibinfo
  {pages} {3865} (\bibinfo {year} {1996})}\BibitemShut {NoStop}%
\bibitem [{\citenamefont {Kaltak}(2015)}]{Kaltak2015}%
  \BibitemOpen
  \bibfield  {author} {\bibinfo {author} {\bibfnamefont {M.}~\bibnamefont
  {Kaltak}},\ }\href {http://othes.univie.ac.at/38099} {\bibfield  {journal}
  {\bibinfo  {journal} {PhD Thesis, University of Vienna}\ } (\bibinfo {year}
  {2015})}\BibitemShut {NoStop}%
\bibitem [{\citenamefont {Souza}\ \emph {et~al.}(2001)\citenamefont {Souza},
  \citenamefont {Marzari},\ and\ \citenamefont
  {Vanderbilt}}]{PhysRevB.65.035109}%
  \BibitemOpen
  \bibfield  {author} {\bibinfo {author} {\bibfnamefont {I.}~\bibnamefont
  {Souza}}, \bibinfo {author} {\bibfnamefont {N.}~\bibnamefont {Marzari}}, \
  and\ \bibinfo {author} {\bibfnamefont {D.}~\bibnamefont {Vanderbilt}},\
  }\href {\doibase 10.1103/PhysRevB.65.035109} {\bibfield  {journal} {\bibinfo
  {journal} {Phys. Rev. B}\ }\textbf {\bibinfo {volume} {65}},\ \bibinfo
  {pages} {035109} (\bibinfo {year} {2001})}\BibitemShut {NoStop}%
\bibitem [{\citenamefont {Gull}\ \emph {et~al.}(2011)\citenamefont {Gull},
  \citenamefont {Millis}, \citenamefont {Lichtenstein}, \citenamefont
  {Rubtsov}, \citenamefont {Troyer},\ and\ \citenamefont
  {Werner}}]{RevModPhys.83.349}%
  \BibitemOpen
  \bibfield  {author} {\bibinfo {author} {\bibfnamefont {E.}~\bibnamefont
  {Gull}}, \bibinfo {author} {\bibfnamefont {A.~J.}\ \bibnamefont {Millis}},
  \bibinfo {author} {\bibfnamefont {A.~I.}\ \bibnamefont {Lichtenstein}},
  \bibinfo {author} {\bibfnamefont {A.~N.}\ \bibnamefont {Rubtsov}}, \bibinfo
  {author} {\bibfnamefont {M.}~\bibnamefont {Troyer}}, \ and\ \bibinfo {author}
  {\bibfnamefont {P.}~\bibnamefont {Werner}},\ }\href {\doibase
  10.1103/RevModPhys.83.349} {\bibfield  {journal} {\bibinfo  {journal} {Rev.
  Mod. Phys.}\ }\textbf {\bibinfo {volume} {83}},\ \bibinfo {pages} {349}
  (\bibinfo {year} {2011})}\BibitemShut {NoStop}%
\end{thebibliography}%

\end{document}